\documentclass[tighten,twocolumn]{aastex631}

\usepackage{CJK}
\usepackage[T1]{fontenc}
\usepackage{gensymb}
\usepackage{graphicx}
\usepackage{subfigure}
\usepackage{xspace}

\newcommand\xmm{\textit{XMM-Newton}\xspace}
\newcommand\nustar{\textit{NuSTAR}\xspace}
\newcommand\srge{\textit{SRG}/eROSITA\xspace}

\shortauthors{Zhang et al.}

\graphicspath{{./}}
\begin{document}
\begin{CJK*}{UTF8}{gbsn}
\title{Modeling Multiple X-Ray Reflection in Super-Eddington Winds}
\author[0000-0001-8701-2116]{Zijian Zhang (张子健)}
\affiliation{Department of Physics, University of Hong Kong, Pokfulam Road, Hong Kong}
\author[0000-0003-4256-7059]{Lars Lund Thomsen}
\affiliation{Department of Physics, University of Hong Kong, Pokfulam Road, Hong Kong}
\author[0000-0002-9589-5235]{Lixin Dai}\email{lixindai@hku.hk}
\affiliation{Department of Physics, University of Hong Kong, Pokfulam Road, Hong Kong}
\author[0000-0002-1510-4860]{Christopher S. Reynolds}
\affiliation{Department of Astronomy, University of Maryland, College Park, MD20742, USA}

\author[0000-0003-3828-2448]{Javier A. Garc\'ia}
\affiliation{X-ray Astrophysics Laboratory, NASA Goddard Space Flight Center, Greenbelt, MD 20771, USA}
\affiliation{Cahill Center for Astronomy and Astrophysics, California Institute of Technology, Pasadena, CA 91125, USA}
\author[0000-0003-0172-0854]{Erin Kara}
\affiliation{MIT Kavli Institute for Astrophysics and Space Research, Massachusetts Institute of Technology, 77 Massachusetts Avenue, Cambridge, MA 02139, USA}
\author[0000-0002-8908-759X]{Riley Connors}
\affiliation{Villanova University, Department of Physics, Villanova, PA 19085, USA}
\author[0000-0003-4127-0739]{Megan Masterson}
\affiliation{MIT Kavli Institute for Astrophysics and Space Research, Massachusetts Institute of Technology, 77 Massachusetts Avenue, Cambridge, MA 02139, USA}

\author[0000-0001-6747-8509]{Yuhan Yao}
\affiliation{Miller Institute for Basic Research in Science, 468 Donner Lab, Berkeley, CA 94720, USA}
\affiliation{Department of Astronomy, University of California, Berkeley, CA 94720, USA}
\author[0000-0003-4583-9048]{Thomas Dauser}
\affiliation{Remeis-Observatory \& ECAP, FAU Erlangen-N\"urnberg, Sternwartstr. 7, 96049 Bamberg, Germany}
\begin{abstract} 
\noindent 
It has been recently discovered that a few super-Eddington sources undergoing black hole super-Eddington accretion exhibit X-ray reflection signatures. In such new systems, one expects that the coronal X-ray emissions are mainly reflected by optically thick super-Eddington winds instead of thin disks. In this paper, we conduct a series of general relativistic ray-tracing and Monte Carlo radiative transfer simulations to model the X-ray reflection signatures, especially the characteristic Fe K$\alpha$ line, produced from super-Eddington accretion flows around non-spinning black holes. In particular, we allow the photons emitted by a lamppost corona to be reflected multiple times in a cone-like funnel surrounded by fast winds. We find that the Fe K$\alpha$ line profile most sensitively depends on the wind kinematics, while its exact shape also depends on the funnel open angle and corona height. Furthermore, very interestingly, we find that the Fe K$\alpha$ line can have a prominent double-peak profile in certain parameter spaces even with a face-on orientation. Moreover, we compare the Fe K$\alpha$ line profiles produced from super-Eddington and thin disks and show that such lines can provide important insights into the understanding of black hole systems undergoing super-Eddington accretion.
\end{abstract}

\section{Introduction} \label{sec: intro}
\noindent Accretion onto black holes (BHs) powers many luminous astronomical systems. For example, supermassive black holes (SMBHs) exist at the centers of most massive galaxies \citep{1995ARA&A..33..581K}, which have grown from seeds through the accretion of gas and stars as well as mergers \citep[e.g.][]{2004MNRAS.351..169M, Hugo21}. In the accretion process, the mass energy of the accreting material can be efficiently transformed into radiation and other forms of energy such as the kinetic energy carried by winds and jets, which can interact with the surrounding gas and provide feedback to the host galaxy \citep{King15}.

There exists a theoretical upper limit of the luminosity produced through spherical accretion onto a BH, which is called the Eddington limit. The Eddington limit can be calculated using the balance of gravity and radiation pressure, and it is given by:
\begin{equation}
    L_{\rm Edd} \approx 1.26\times 10^{38}\left(\frac{M_{\rm BH}}{M_{\sun}}\right)\ \rm{erg\ s^{-1}}
\end{equation}
for fully ionized hydrogen gas. Here $M_{\rm BH}$ and $M_\sun$ are the black hole mass and the solar mass separately.
The corresponding  accretion rate is called the Eddington accretion rate:
\begin{equation}    
    \dot M_{\rm Edd} = \frac{L_{\rm Edd}}{\eta c^2},
\end{equation}
where $\eta$ is the radiative efficiency, and we usually take a nominal value of $10\%$.

Super-Eddington accretion, in which gas is supplied to the black hole at rates exceeding $\dot M_{\rm Edd}$ (which can be partly enabled by the departure from spherical symmetry in disk-like systems), happens in many important BH systems. Very luminous quasars found in high-redshift galaxies \citep[e.g.][]{1999ApJ...526L..57F, Wu15, 2021ApJ...907L...1W} likely have gone through episodes of super-Eddington accretion to build up their masses \citep{Volonteri15}. Also, transient super-Eddington accretion phenomenon can occur in tidal disruption events (TDEs), in which stars that get too close to SMBHs are disrupted and their debris is accreted onto the BH \citep{Rees88,2021ARA&A..59...21G, Dai21}. In the stellar-mass regime, a few black hole binaries (BHBs) are observed to be super-Eddington in the soft state \citep[e.g.][]{2017MNRAS.471.1797M,Negoro21b,Connors21, Prabhakar23,Jin24,Zhao24}. Furthermore, there is more and more observational evidence that ultraluminous X-ray sources (ULXs) are powered by neutron stars and stellar-mass black holes at accretion rates highly above $\dot M_{\rm Edd}$ \citep{2017ARA&A..55..303K}.

Accretion flows in the super-Eddington regime have structures and physics that are distinct from standard thin disk models \citep{SS73, Novikov73}. Super-Eddington accretion flows are geometrically and optically thick, and photons in the inner disks can be easily advected into the black hole leading to low radiative efficiency \citep{Begelman78, 1980ApJ...242..772A}. Recently, a lot of insight into super-Eddington accretion physics has been gained through performing novel general relativistic radiation magnetohydrodynamic (GRRMHD) simulations \citep[e.g.][]{2014MNRAS.441.3177M, Jiang14, Jiang19, 2015MNRAS.447...49S, 2016MNRAS.456.3929S, Dai18, Thomsen22b}. These simulations reveal the critical aspect that fast and optically thick winds are launched from super-Eddington accretion flows due to the large radiation and magnetic pressure.  Such winds have been observed in several TDEs \citep{Kara16a, 2018MNRAS.474.3593K,2016ApJ...819L..25A, 2017NatAs...1E..33L,2019ApJ...879..119H}. Simulations reveal that super-Eddington winds have anisotropic structures as illustrated in Fig.~\ref{schematic}  (see \citealt{Dai18} for details). They are denser and slower ($v<0.1c$) at larger inclinations while being dilute and ultrafast ($v\gtrsim 0.1c$) close to the pole. This wind structure naturally produces an optically thin ``funnel'' around the pole, through which beamed radiation can easily leak out. 

In recent years, a technique called X-ray reflection spectroscopy has been developed, which can be used to map out the structure of the innermost accretion flow in active galactic nuclei (AGNs) and X-ray binaries (XRBs) \citep{1989MNRAS.238..729F, Dauser10, 2021ARA&A..59..117R}. The X-ray emissions from such systems contain a non-thermal power-law component produced by the corona, which is generally believed to be a region where high-energy electrons up-scatter the thermal photons from the disk \citep{2021iSci...24j2557C}. The emission from the corona can irradiate the accretion disk and produce a reflection spectrum, with its most significant signature being the fluorescent Fe K$\alpha$ emission lines. The line profiles are shaped by both general relativistic effects and Doppler motion of the emitting gas, which can be used to probe black hole spin and corona-disk geometry \citep{Wilkins12, 2021ARA&A..59..117R}. As photons along different paths have different light travel time, one can also use the lag between the corona direct emission and reflected emission to carry out reverberation studies and obtain information about the disk, corona, and black hole\citep[e.g.][]{2012MNRAS.422..129Z, 2014MNRAS.438.2980C, Uttley14, Kara16b,Kara19, 2021Natur.595..657W}.

While the study of X-ray reflection has so far focused on sub-Eddington systems, recently, a few super-Eddington black hole systems have been observed to exhibit X-ray reflection features. In such systems, it is expected that the corona photons are mostly reflected in the optically thick wind and the reflection signatures therefore should deviate from those produced in thin disks.
For example, \citet{Kara16a} observed the blueshifted Fe K$\alpha$ lag in a jetted TDE Swift J1644+57 \citep{Burrows11, Zauderer11, Bloom11}. On the theoretical front, \citet{2019MNRAS.482.5316M} used a simple funnel geometry surrounded by wind and considered scattering and absorption in the wind and showed that the lag spectra are consistent with those observed from two AGNs which are likely super-Eddington. \citet{Thomsen19, Thomsen22a}, instead, started from a super-Eddington accretion flow from a previous state-of-the-art numerical simulation \citep{Dai18} and computed the energy and lag spectrum of the Fe K$\alpha$ line. They employed rigorous GR ray-tracing calculations and showed that indeed the reflection surface for the corona photons resides mostly in the fast wind launched in super-Eddington accretion, and one expects to see the reflection signatures only when looking down at the funnel. Furthermore, they showed that the Fe line energy and lag spectral profiles are more blueshifted and symmetric compared to the thin disk case, and the lags are generally shorter because the reflection is produced by the fast wind instead of a thin disk. This result is supported by \citet{Masterson22}, in which a blueshifted Gaussian-like 1\,keV (possibly O) line is detected during the super-Eddington phase of a changing-look AGN.

These emerging observations show that more extensive theoretical modelings of X-ray reflection in the context of super-Eddington accretion are urgently needed. For example, previous literature as mentioned above has not yet studied how the reflection spectrum is affected by funnel open-angle and wind acceleration profiles, which are likely linked to the BH accretion rate and spin \citep{2016MNRAS.456.3929S}. Also, they have missed an important component, which is multiple reflections of the X-ray photons across the funnel (as illustrated in Fig.~\ref{schematic}). Multiple reflection, or sometimes phrased as returning radiation, has recently been noticed to produce a non-negligible effect on the reflection spectra from thin accretion disks \citep{Wilkins20, Dauser22, Mirzaev24}. For super-Eddington systems with a narrow funnel geometry, we expect that a higher fraction of the photons can go through multiple reflections before escaping, and therefore this effect is worthy of a detailed study.
 
In this work, we conduct a systematic study of the fluorescent Fe K$\alpha$ line profiles in super-Eddington accretion flows. We focus on two aspects missed by previous studies: 1) How does the multiple reflection of the X-ray photons across the funnel affect the observed signatures? 2) How can we use the profiles of Fe K$\alpha$ lines to probe key parameters such as the funnel open angle, the terminal velocity and the acceleration profile of the wind, and the height of the corona?  The paper is structured as follows. We introduce our model setup and methodology in Section~\ref{sec: method}. We show the results of the Fe line spectral features in Section~\ref{sec:spec}, where one can see that multiple reflections can produce a prominent double-peak feature in the energy spectrum (Section~\ref{feature}). We investigate how various physical parameters related to the funnel geometry and wind kinematics affect the line profiles in Section~\ref{sec:parameters}. We show the parameter space preferred for producing the double-peak line feature in Section~\ref{sec: 2D} and compare the morphology of the Fe K$\alpha$ lines produced from thin disks and super-Eddington accretion flows in Section~\ref{morpho}. We apply our model to the spectrum observed from a Galactic BHB 4U1543--47 in a super-Eddington outburst phase in Section \ref{sec:fit}. We summarize and discuss future applications of our work in Section~\ref{sec: summary}.

\section{Methods} \label{sec: method}

\subsection{Setup of the Super-Eddington Funnel-Wind-Corona Model}
\begin{figure}[!htb]
    \centering
    \includegraphics[width=1\linewidth]{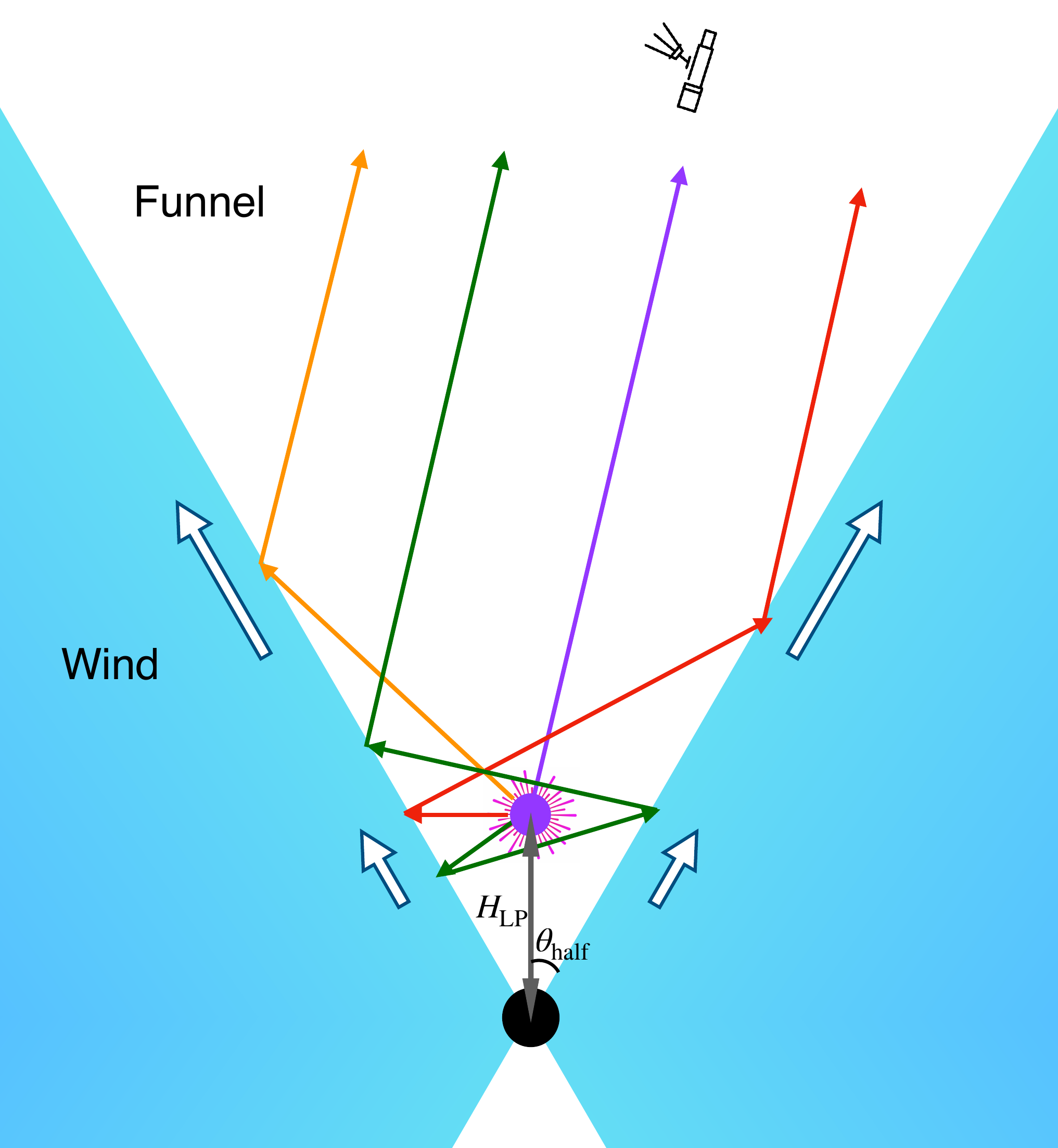}
    \caption{A schematic diagram showing how the photons emitted by a BH corona are reflected  multiple times inside the super-Eddington funnel surrounded by optically thick wind. }
    \label{schematic}
\end{figure}

\noindent We adopt a simple model as illustrated in Fig.~\ref{schematic} to represent the reflection geometry in super-Eddington accretion flows. We set up a conical funnel without gas inside. The half-open angle of the funnel is $\theta$.  The funnel wall is cut off at a radial distance of $400\,R_g$ from the BH. For simplicity, we assume that the corona is a point-like source and adopt a lamppost model, although we recognize that BH coronae likely have more complicated geometry and structure  \citep{Kara19, WangJY21}. The corona is placed along the polar axis with a height of $h_{\rm lp}$ above the BH. 

\begin{figure}[!htb]
    \centering
    \includegraphics[width=1\linewidth]{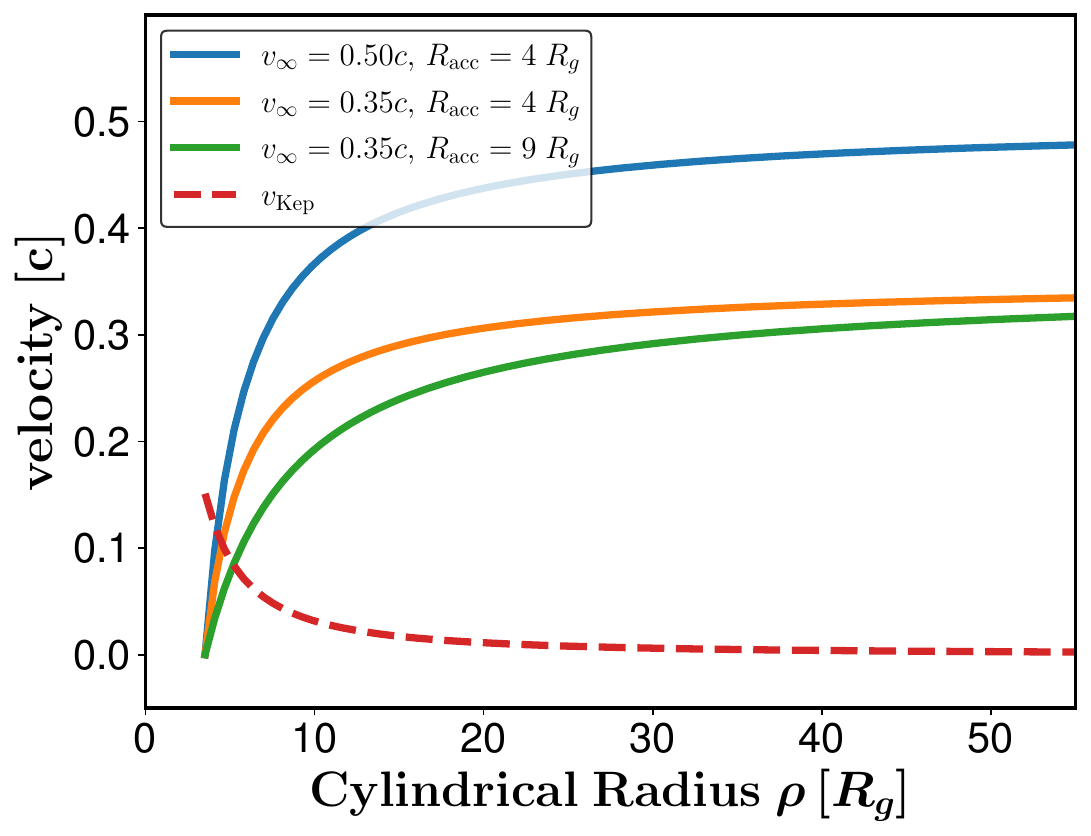}
 \caption{The wind radial velocities with different $v_{\infty}$ and $R_{\rm acc}$ (solid curves)
 （with $\theta_{\rm half} = \pi/5$） in comparison to the Keplerian rotation velocity (dashed red curve). The maximum radial velocity of the wind is determined by $v_{\infty}$, and $R_{\rm acc}$ controls the wind acceleration profile. The wind radial velocity dominates over rotation velocity $\rho \gtrsim 5\,R_g$.}
    \label{speed}
\end{figure}

The funnel is surrounded by a layer of optically thick wind, which is assumed to be launched from the inner disk and accelerates radially until reaching a terminal velocity $(v_\infty)$. The wind radial velocity profile is described by the extended CAK law \citep{1975ApJ...195..157C,Thomsen19,2022MNRAS.510.5426P}:
\begin{equation}{\label{cak}}
    v = v_0+(v_{\infty}-v_0)\frac{(l/R_{\rm acc})^{\alpha}}{(l/R_{\rm acc})^{\alpha}+1}.
\end{equation}
Here the wind is launched with a zero velocity $v_0 = 0$ and accelerates to the terminal velocity $v_{\infty}$ faraway. $l = r-r_{\rm base}$ is the distance from the wind-launching point, which is located at $r_{\rm base}$ away from the BH. We assume that $r_{\rm base} = R_{\rm ISCO}$, the radius of the innermost stable circular orbit, and $R_{\rm ISCO}=6\,R_g$ around a non-spinning black hole (with $R_g\equiv GM_{\rm BH}/c^2$ being the BH gravitational radius).  $R_{\rm acc}$ is the wind acceleration radius. $\alpha$ is the wind acceleration index for which we fix to be 1. The velocity profile is shown in Fig.~\ref{speed} for a few characteristic values of $v_{\infty}$ and $R_{\rm acc}$. One can see that the wind accelerates faster when $R_{\rm acc}$ is smaller. Also, in super-Eddington accretion, the radial velocity of the wind usually largely exceeds the Keplerian velocity. Therefore, we ignore the effect of wind rotation in this work. 

\begin{deluxetable}{cccc}

\tablecaption{Free parameters and their default value in our model}
\label{parameters}
\tablehead{\colhead{parameter} & \colhead{default} & \colhead{range} & \colhead{interval}} 
\startdata
$v_{\infty}$        & $0.35c$   & $[0.0,\ 0.5c]$            &$0.05c$\\
$R_{\rm acc}$       & $4\,R_g$   & $[2\,R_g,\ 20\,R_g]$       &$2\ R_g$\\
$H_{\rm LP}$        & $10\,R_g$  & $[5\,R_g,\ 30\,R_g]$       &$2.5\,R_g$\\
$\theta_{\rm half}$ & $\pi/5$   & $[\pi/10,\ \pi/4]$        &$\pi/60$\\
$i$                 & 0         & $[0,\ \theta_{\rm half}]$ &$\theta_{\rm half}/10$\\
\enddata
\end{deluxetable}

We summarize the free parameters of our model in Table~\ref{parameters}: wind terminal velocity $v_{\infty}$, wind acceleration radius $R_{\rm acc}$, height of the corona $H_{\rm LP}$, half open angle of the funnel $\theta_{\rm half}$ and observer inclination angle $i$. Their default values and ranges are also listed.

\subsection{General Relativistic Ray-Tracing and Monte-Carlo Simulation}
\noindent We adopt a forward ray-tracing method and have all incident photons emitted from the lamppost corona with a power-law distribution of energy \citep[e.g.][]{1991MNRAS.249..352G},
\begin{equation}
    N(E)=AE^{-\Gamma}\ {\rm photons\,keV^{-1}},
\end{equation}
where we take $\Gamma \equiv 2$ in this work.

We calculate the photon trajectory inside the funnel using a rigorous GR ray-tracing code \citep{Dai10, Thomsen19, Thomsen22a}. Here we assume that the BH spin is set to 0 and use the Schwarzschild metric. If the photon crosses the funnel wall and enters the wind, we take a Monte Carlo approach to calculate its propagation inside the wind and possible interaction with the gas.  To simplify the computation, the step size of the photon $({\rm d}l)$ is fixed to be 0.1 $r_{\tau}$, where $r_{\tau}$ is the mean free path of a photon, which is set to be $1\,R_g$ by assuming a constant electron number density. As the photon step size inside the wind is small \citep{Thomsen19}, we ignore GR effects and use a Newtonian calculation for the photon propagation. After each step, the photon can be scattered by gas, or be absorbed by an Fe ion (which then possibly produces an Fe K line photon), or go through no interaction at all. For the last scenario, the photon retains its original direction of motion. On the other hand, if an interaction happens, the direction of the scattered or re-emitted photon will have isotropic distribution in the emitting gas frame, and the exact direction is randomly chosen each time. If the scattered or re-emitted photon enters the funnel, we follow its trajectory with the GR ray-tracing code again and this process repeats. 
 
The details of these radiative transfer processes inside the wind are described below:

1) Scattering: The scattering of the photon is assumed to be elastic for simplicity given most coronal X-ray photons have relatively low energy compared to the rest mass energy of an electron. Therefore, after scattering the energy of the photon does not change while its new direction is isotropically distributed in the gas frame. 
(It is noted that the consideration of inelastic and anisotropic scattering would merit a more accurate calculation. We have conducted tests for the same model using elastic scattering and inelastic scattering in Appendix \ref{elastic}, and found that the two treatments give similar first-order results for the energy range considered in this work.) As the optical depth for the medium traveled through in one step is $\tau=0.1$, therefore, the escape probability is $e^{-0.1}$ and the probability for scattering to happen is $1-e^{-0.1}\sim 0.1$. In practice, most scattering events happen within $\tau=3$.

2) Iron K absorption/line emission: In this study, we only consider the absorption of the X-ray photon by Fe. To be more exact, the electron at the K-shell of Fe can be ionized by the energetic coronal photon, and then Fe K lines can be produced when the L-shell electrons transition to the K-shell. The exact energy of the Fe K line depends on the ionization stage of the Fe ion, which is described by the ionization parameter $\xi(r)=4\pi F_X(r)/n(r)$, where $F_X(r)$ is X-ray flux at radius $r$, and $n(r)$ is the electron number density \citep{1997ApJ...488..109R}. The super-Eddington winds are dilute, so the ionization parameter is typically high \citep{1982JPCRD..11..135C, Thomsen19, Thomsen22a}. For this study, we mainly consider the 6.7\,keV Fe K$\alpha$ line produced by He-like \ion{Fe}{25} with an 8.8\,keV ionization energy, which requires the recombination of a free electron to be produced. However, here we still call it a fluorescent line for consistency with literature. In our Monte-Carlo calculation, if a photon has energy higher than 8.8\,keV, it can be absorbed by the wind and possibly re-emits a 6.7\,keV Fe K$\alpha$ photon. The probability for the photon to be absorbed is $1-e^{-\tau_{\rm abs}}$, where $\tau_{\rm abs} = \sigma_{\rm abs} n_{\rm Fe} {\rm d}l$. The cross-section $(\sigma_{\rm abs})$ of the photoionization is taken from \citet{1995A&AS..109..125V} and the iron abundance is fixed to the solar value from \citet{1998SSRv...85..161G}. The probability for recombination to happen after the photon absorption is the so-called effective yield. We take an effective yield of 0.5 following \citet{KK87} for an order-of-magnitude calculation.

The X-ray photons can travel out of the funnel and re-enter the wind multiple times due to the narrow funnel geometry. For example, a photon emitted by the corona can travel through the funnel, enter the wind (first reflection event) where it is scattered by the wind to enter the funnel again, and then enter the wind again (second reflection event) and this time a fluorescence event happens and an Fe line photon is produced, and this Fe line photon can enter the wind one more time (third reflection event) and gets scattered, and then this process continues. We trace all the photons until they either reach the far-away observer placed at 400\,$R_g$ away from the corona, or enter the black hole horizon. We also stop tracing the photons that have gone scattering inside the wind for more than 10 times in total during the reflection events.

\section{Energy spectrum}
\label{sec:spec}
\noindent We first introduce the composition of the reflection energy spectrum. A characteristic spectrum is shown in Fig.~\ref{FeLine}\,(a), which is obtained using the fiducial model with the following default parameters: $v_\infty = 0.35c,\,R_{\rm acc} = 4\,R_g,\,H_{\rm LP}=10\,R_g,\,\theta_{\rm half} = \pi/5$ and the observer inclination $i=0$ (face-on). 

In the lower energy range ($\sim\!2-5\,\rm keV$), photons can only get scattered, so the reflection spectrum has the same power-law form as the coronal spectrum. Between $\sim\!5-9\,\rm keV$, the Fe K$\alpha$ line dominates. Above this energy, the Fe K absorption edge appears, as photons with energy higher than a threshold ($8.8\,\rm keV$ for our model) can be absorbed. Since the absorption cross-section decreases as the energy increases above this threshold, the spectrum gradually approaches the original power-law at higher energies. We also further decompose the overall spectrum into three components in Fig.~\ref{FeLine}\,(a) -- the contribution from the primary reflection (photons gone through only one reflection event by the wind, orange line), secondary reflection (photons reflected twice by the wind, blue line) and the rest (higher-order reflections, green line).

We focus on analyzing the properties of the most significant feature of the energy spectrum -- the broad Fe K$\alpha$ line -- in this article. In this Section, we will first show the detailed Fe K$\alpha$ line profile and composition in Section~\ref{feature}, followed by a discussion on how the Fe line profile depends on various physical parameters in Section~\ref{sec:parameters}. We will highlight the parameter space in which the Fe line exhibits a unique double-peak feature in Section~\ref{sec: 2D}. We compare the double-peak Fe line profile from thin-disks and super-Eddington accretion flows and show their morphological differences in Section~\ref{morpho}. Finally, we briefly discuss the properties of the absorption edge in Appendix \ref{absorption}.

\subsection{The composition of the Fe K$\alpha$ line profile with a possible double peak feature}\label{feature}

\begin{figure*}[!htb]
    \centering
    \includegraphics[width=0.95\linewidth]{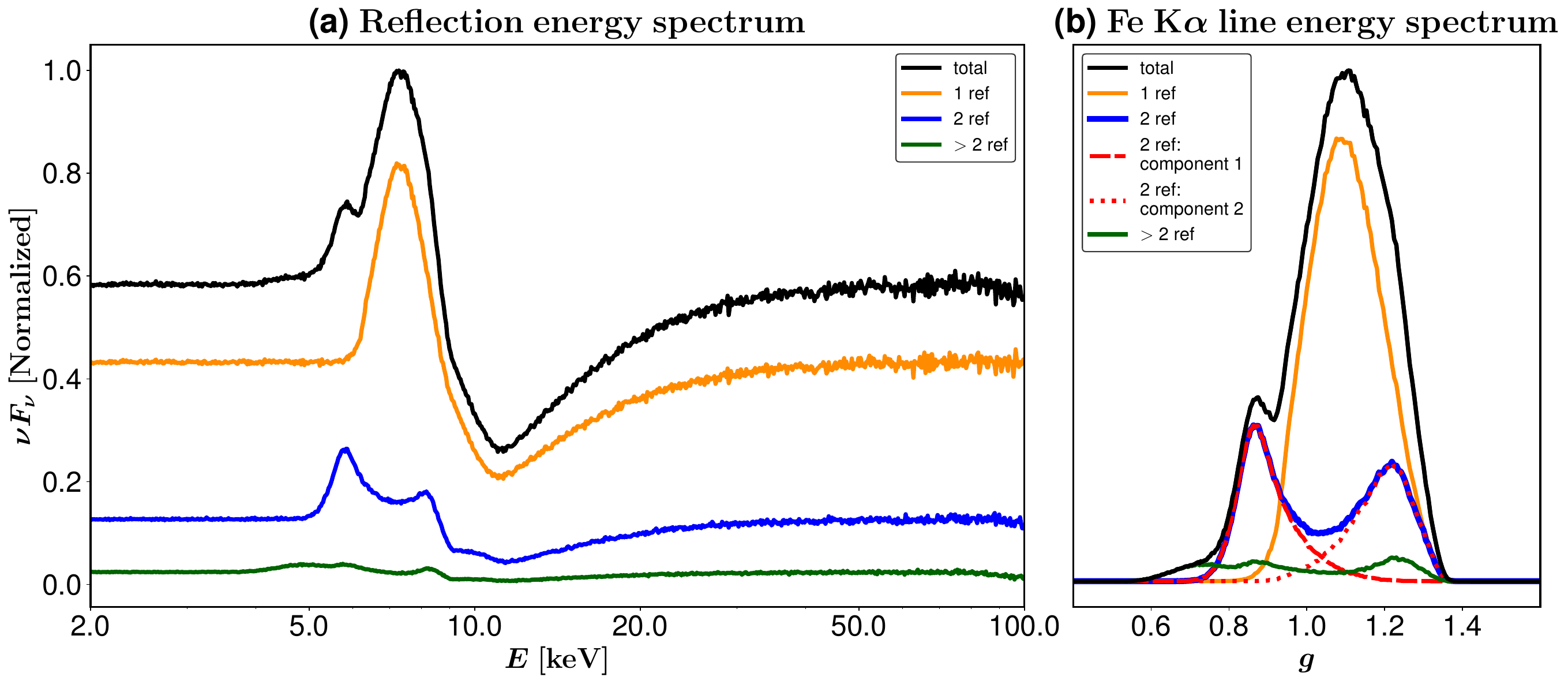}
     \caption{ The overall reflection energy spectrum (left, panel a) and the Fe K$\alpha$ line energy spectrum (right, panel b). The spectra in both panels are produced using the fiducial model with the default parameters: $v_\infty = 0.35c$, $R_{\rm acc} = 4\,R_g$, $H_{\rm LP}=10\,R_g$, $\theta_{\rm half} = \pi/5$ and $i=0$ (face-on). The x-axis in panel (a) is the observed energy of the photon (in keV), while that in panel (b) is the energy shift of the photon $g = E_{\rm obs}/E_{\rm emit}$, where $E_{\rm emit} = 6.7\, {\rm keV}$. The y-axes in both panels are the normalized fluxes. The black solid line denotes the total reflection energy spectrum, with colored curves showing contributions from reflections of different orders: orange --  photons that have undergone one reflection; blue -- photons that have undergone two reflections; green -- photons that have undergone three or more reflections. For the blue curve in panel (b), we further decompose it into two components: red dotted curve -- coronal photons are only scattered in the first reflection and then the scattered photons produce Fe K$\alpha$ photons in the second reflection; red dashed line -- coronal photons have already produced Fe K$\alpha$ photons in the first reflection and then the Fe K$\alpha$ photons are scattered in the second reflection.}
    \label{FeLine}
\end{figure*}

\noindent The Fe line profile for our fiducial model is shown in Fig.~\ref{FeLine}\,(b), where different reflection components contributing to the overall spectrum are shown.

The primary Fe K$\alpha$ reflection energy spectrum (orange line) is composed of the Fe photons that are produced as the coronal photons irradiate the wind, which is capable of escaping through the funnel right afterward. 
One can see that this line profile is blueshifted, since the photons are reflected by the wind which travels towards the observer. Also, the line is broadened because the wind has a certain acceleration profile and the projected velocity along the line of sight is different. The line has a relatively symmetric shape with respect to the peak, which can be a criterion used to differentiate lines produced from super-Eddington accretion flow or from thin disks (the latter usually have more skewed shapes). Such line features have been previously investigated by \citet{Thomsen19, Thomsen22a}, who obtained consistent results, although they have used the geometry of a super-Eddington accretion flow from simulations while we adopt a simple cone geometry for the funnel. 

Very interestingly, the Fe line produced from the secondary reflection (blue line) has two notably different components. The first component (red dashed line) is formed by the Fe line photons which are already produced in the first reflection and then re-enter the wind and get scattered before escaping through the funnel. In this scenario, as the winds on different sides of the funnel travel away from each other, the Fe line photons generally lose energy in the observer's frame after the reflection, which therefore introduces a redshift to the spectral peak.
On the other hand, the second component (red dotted line) is produced when coronal photons are scattered by the wind during the first reflection and then re-enter the wind and produce the Fe K line during the second reflection before escaping. 
As the second reflection usually happens at a relatively outer part of the funnel, where the wind has accelerated to a higher velocity, the Fe line produced in this scenario usually has an even larger blueshift compared to the Fe line produced from the primary reflection.

Similarly, the third reflection should have three spectral peaks and so on, although these high-order components are too weak to produce observable effects. We plot all of the higher-order Fe reflection spectrum components together (green line). 

In summary, the inclusion of multiple photon reflections into our model leads to two important features in the Fe line energy spectrum. First, the overall line profile is further broadened. Second and more interestingly, a secondary redshifted peak can be produced in certain parameter space, more details of which will be shown in Section~\ref{sec:parameters} and \ref{sec: 2D}.

\begin{figure*}[!htb]
    \centering
    \includegraphics[width=0.9\linewidth]{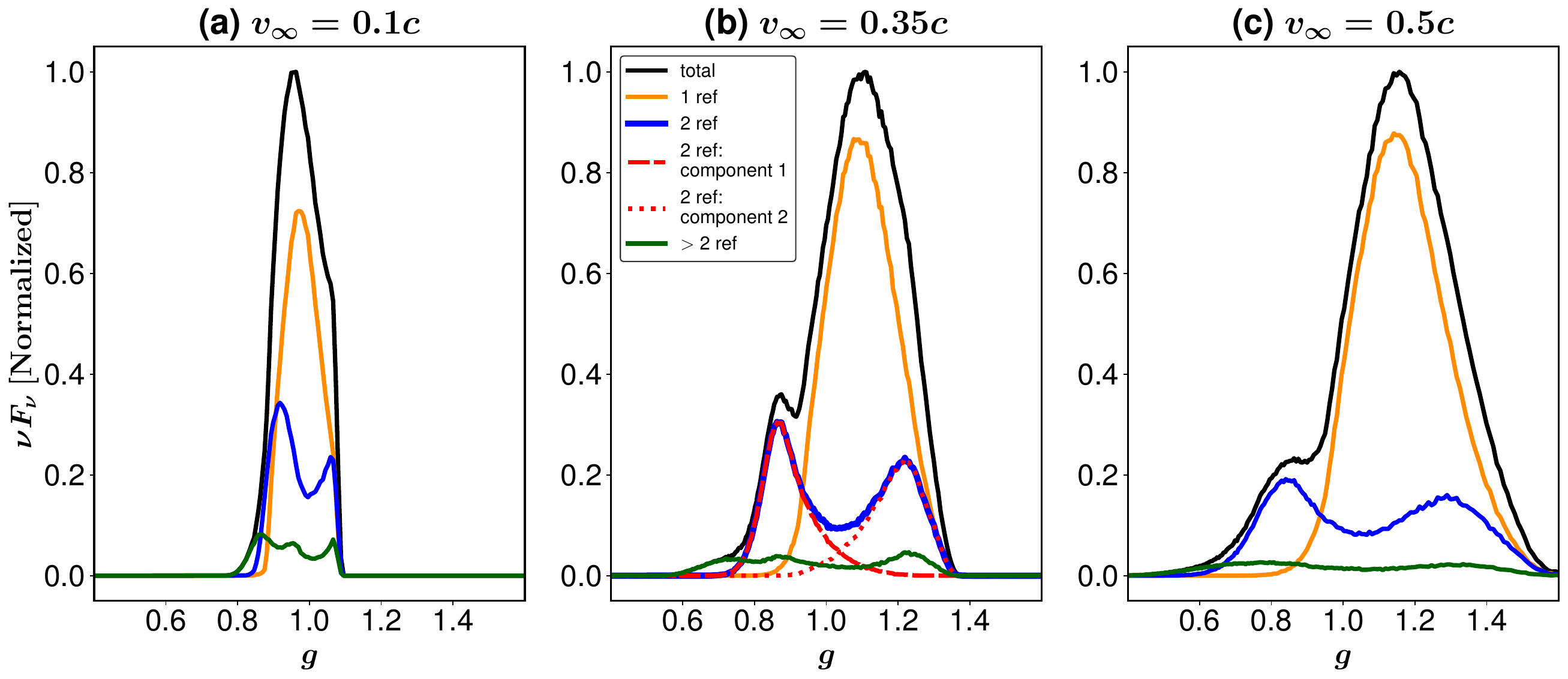}
     \caption{Fe K$\alpha$ line spectrum dependence on the wind terminal velocity $v_\infty$. The color scheme and line styles are the same as in Fig.~\ref{FeLine}.
     Panel (b) uses the default parameter set (the same as in Fig.~\ref{FeLine}), while the other two panels have different $v_\infty$. One can see that increasing $v_\infty$ increases both the blueshift of the primary peak and the redshift of the secondary peak, as well as increases the separation but decreases the flux ratio between the secondary and primary peaks. For very low $v_\infty$, the two peaks merge into one.}
    \label{vinf}
\end{figure*}

\subsection{Fe K$\alpha$ line profile dependence on physical parameters}\label{sec:parameters}

\noindent  We next investigate how the Fe K$\alpha$ energy spectrum profile depends on the four physical parameters of our model, namely, the wind terminal velocity $v_{\infty}$, wind acceleration radius $R_{\rm acc}$, height of the corona $H_{\rm LP}$, half open angle of the funnel $\theta_{\rm half}$, as well as the observer inclination angle $i$. 
For all figures shown in this subsection, the default parameters as listed in Table~\ref{parameters} are used unless specified otherwise.

\subsubsection{Dependence on wind kinematics}
\label{sec:kinematics}
\noindent We first check how the Fe line profile depends on the kinematics of the disk wind, which is described by two free parameters in our model (Eq. \ref{cak}) -- $v_{\infty}$ (terminal velocity) and $r_{\rm acc}$ (acceleration radius).

The dependence of the Fe line spectrum on $v_{\infty}$ is shown in Fig.~\ref{vinf}. As $v_{\infty}$ increases, wind moves faster all the way along the funnel wall, and the wind speed has a larger degree of variation in the inner funnel region (which can be seen in Fig.~\ref{speed}). Therefore, a larger $v_{\infty}$ increases both the blueshift and the width of the primary peak.  $v_{\infty}$ also affects the secondary peak as follows. 
As $v_{\infty}$ increases, the photons produced in the first reflection will be more relativistically beamed and then more likely to escape through the funnel after a single reflection. Therefore, we see that increasing $v_{\infty}$ generally reduces the flux ratio between the secondary peak and the primary peak. Also, since the winds on different sides of the funnel travel away from each other at larger speeds, the Fe line photons produced in the first reflection generally lose more energy when scattered by the wind again as described in Section~\ref{feature}, which gives a larger redshift to the secondary peak. As a result of a more redshifted secondary peak and a more blueshifted primary peak, the separation between the two peaks also increases with $v_{\infty}$. In contrast, when $v_{\infty}$ is too low, the two peaks merge together and the double-peak feature disappears, as shown in Fig.~\ref{vinf}(a).

\begin{figure*}[!htb]
    \centering
    \includegraphics[width=0.9\linewidth]{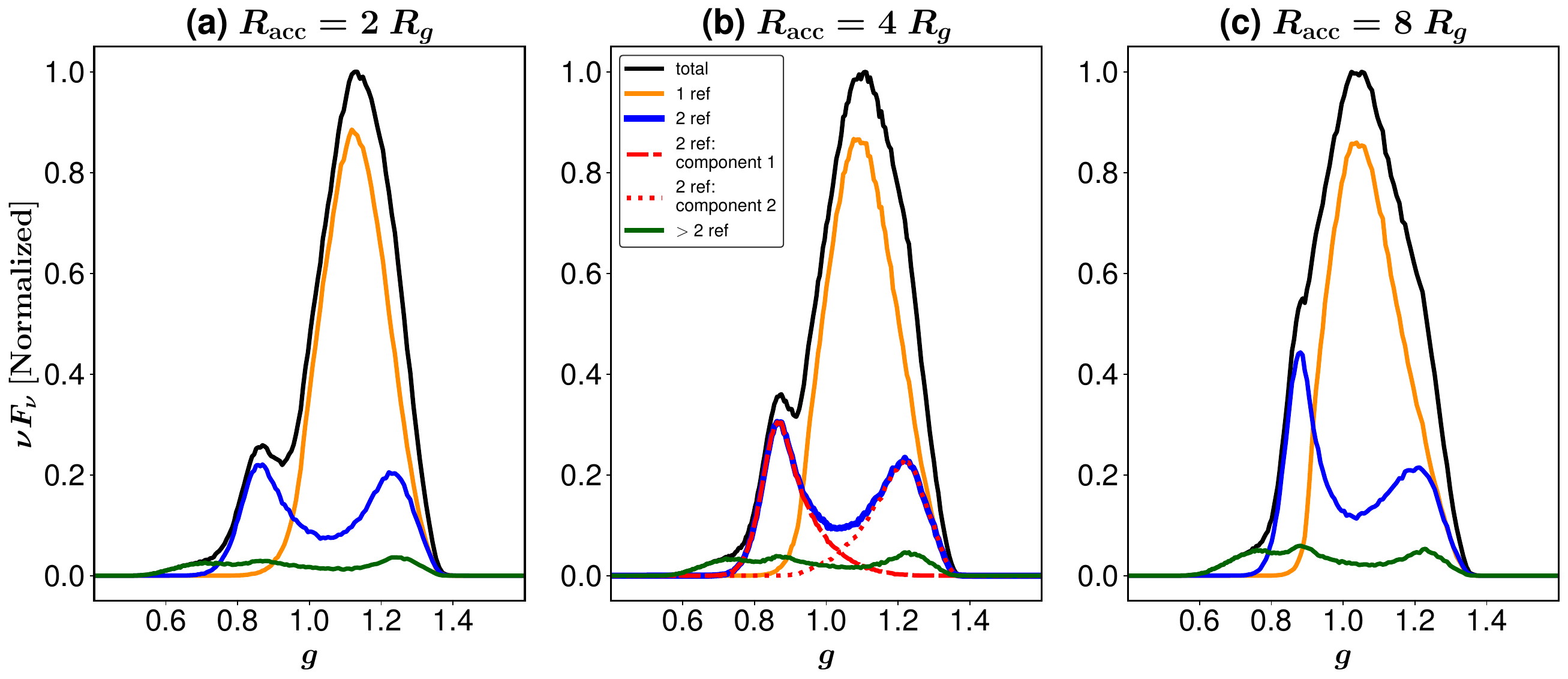}
    \caption{Fe K$\alpha$ line spectrum dependence on the wind acceleration radius $R_{\rm acc}$. The color scheme and line styles are the same as in Fig.~\ref{FeLine}.
     Panel (b) uses the default parameter set (the same as in Fig.~\ref{FeLine}), while the other two panels have different $R_{\rm acc}$. As $R_{\rm acc}$ increases, the wind has a slower acceleration, so increasing $R_{\rm acc}$ produces a similar effect as decreasing $v_\infty$.}
    \label{Racc}
\end{figure*}

The dependence of the Fe line spectrum on $R_{\rm acc}$ is shown in Fig.~\ref{Racc}. A smaller $R_{\rm acc}$ gives faster acceleration of the wind and overall larger wind velocity along the funnel. Therefore, decreasing $R_{\rm acc}$ produces a similar effect on the line spectrum as increasing $v_{\infty}$, i.e.,  a larger blueshift for the primary peak, a larger redshift for the secondary peak, and a smaller flux ratio between the two peaks. Reversely, a slower acceleration profile means the two peaks are more likely to merge into one. While changing  $R_{\rm acc}$ or $v_{\infty}$ can produce similar effects for the peaks, the blue tail of the primary peak is mainly affected by $v_{\infty}$.

\subsubsection{Dependence on funnel geometry and corona height}
\label{sec:kinematics}

\begin{figure*}[!htb]
    \centering
    \includegraphics[width=0.9\linewidth]{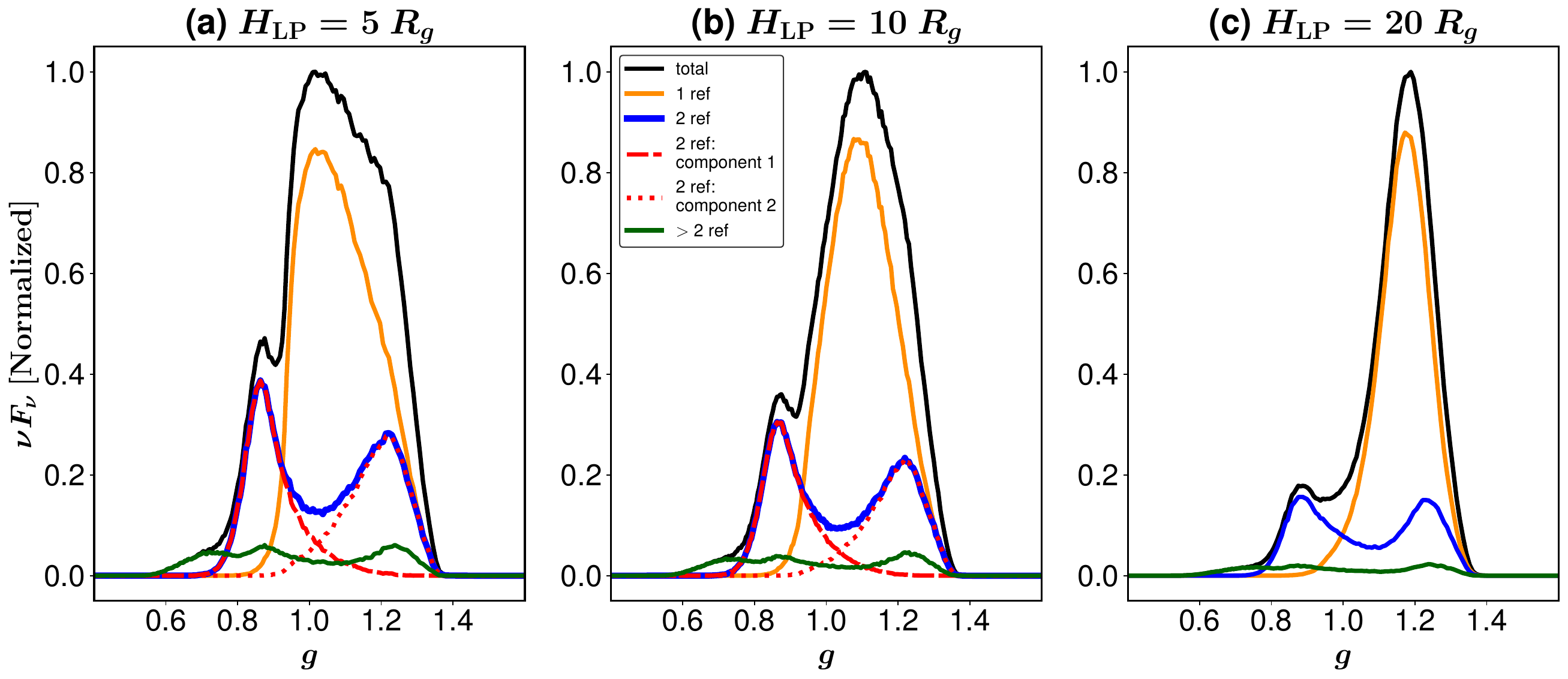}
     \caption{Fe K$\alpha$ line spectrum dependence on the height of corona $H_{\rm LP}$. The color scheme and line styles are the same as in Fig.~\ref{FeLine}.
     Panel (b) uses the default parameter set (the same as in Fig.~\ref{FeLine}), while the other two panels have different $H_{\rm LP}$ as indicated in the title. A higher corona increases the blueshift of the primary peak and reduces the flux ratio between the secondary and primary peaks.} 
    \label{h}
\end{figure*}

\noindent The Fe line spectrum is also affected by two more physical parameters: the height of the lamppost corona $h_{\rm LP}$ and the half-open angle of the funnel $\theta_{\rm half}$.

We first investigate the dependence on $h_{\rm LP}$. Due to the narrow funnel geometry, the wind at a similar height as the corona receives more illumination. Therefore, when a corona is placed higher above the black hole, less photons will be received and reflected in the lower funnel region. This produces two main effects, which can be seen in Fig.~\ref{h}: 1) A higher corona makes it easier for photons to escape after one reflection, which reduces the fluxes produced from multiple reflections. 2) A higher corona illuminates the higher part of the funnel more, where the wind has accelerated to a faster speed, so the primary peak has a larger blueshift, and the separation between the primary and secondary peaks increases. 
 
\begin{figure*}[!htb]
    \centering
    \includegraphics[width=0.9\linewidth]{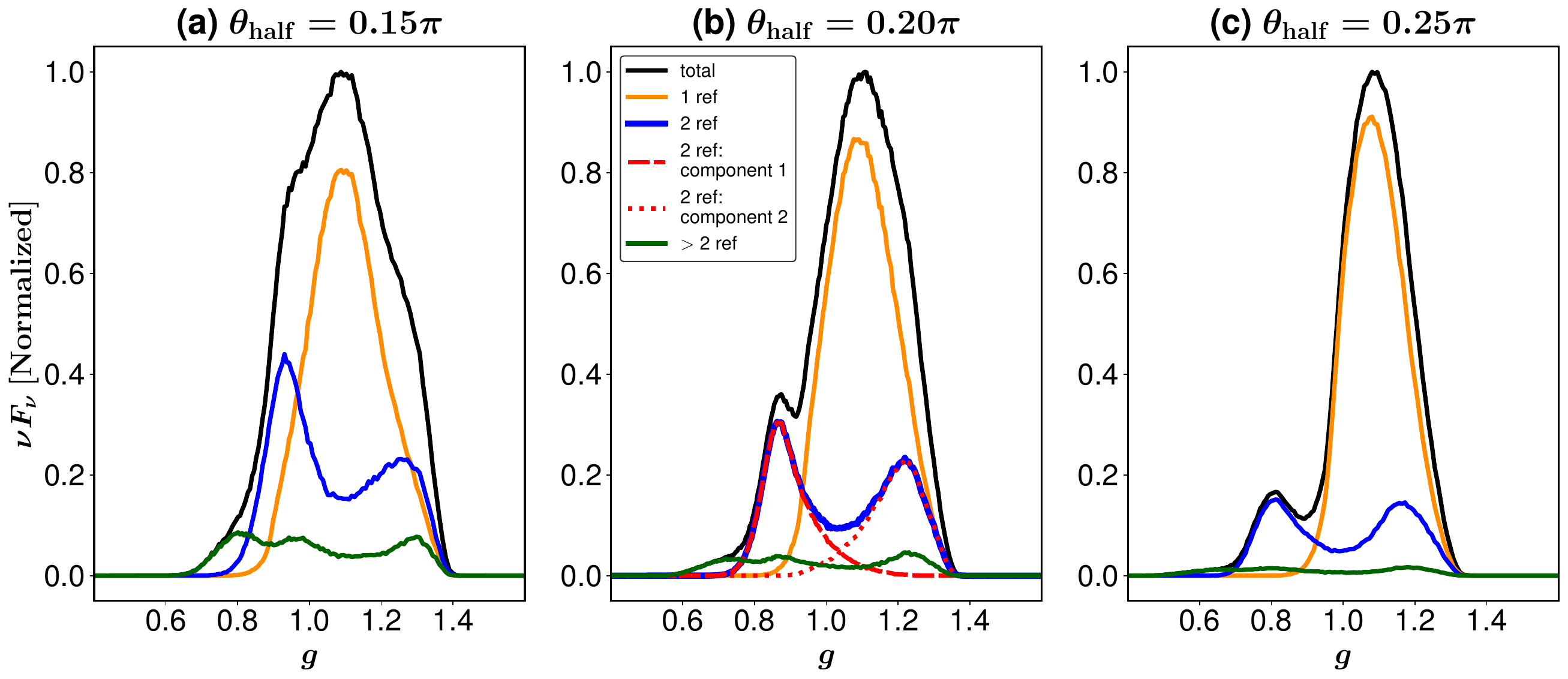}
    \caption{Fe K$\alpha$ line spectrum dependence on the half-open angle $\theta_{\rm half}$. 
    The color scheme and line styles are the same as in Fig.~\ref{FeLine}.  Panel (b) uses the default parameter set (the same as in Fig.~\ref{FeLine}), while the other two panels have different $\theta_{\rm half}$. 
    As  $\theta_{\rm half}$ increases,  the flux ratio between the secondary and primary peaks decreases, and the separation between the two peaks increases. }
    \label{open}
\end{figure*}

Next, we check how the dependence on $\theta_{\rm half}$. As shown in Fig.~\ref{open}, as $\theta_{\rm half}$ decreases, the 
secondary peak is less redshifted. This is because the relative velocity between the winds on different sides of the funnel decreases as the funnel becomes narrower. On the other hand, there is no obvious change for the primary peak, since the emissivity for the primary peak barely depends on $\theta_{\rm half}$. Furthermore, the flux ratio between the secondary and primary peaks decreases with increasing $\theta_{\rm half}$, since a wider funnel makes it easier for photons to escape after the first reflection.

\subsubsection{Dependence on observer inclination}
\label{sec:inclination}
\noindent Observers viewing a super-Eddington accretion flow will only be able to see the emissions from the corona and its reflection when looking into the optically thin funnel, which is typically quite narrow as revealed by recent advanced simulations \citep[e.g.][]{Dai18, Thomsen19, Jiang19}. Therefore, one can expect that the spectral dependence on inclination should be weaker compared to the thin-disk scenario, for which $i$ can vary from $0^\circ$ to $90^\circ$.

\begin{figure}[!htb]
    \centering
    \includegraphics[width=1.0\linewidth]{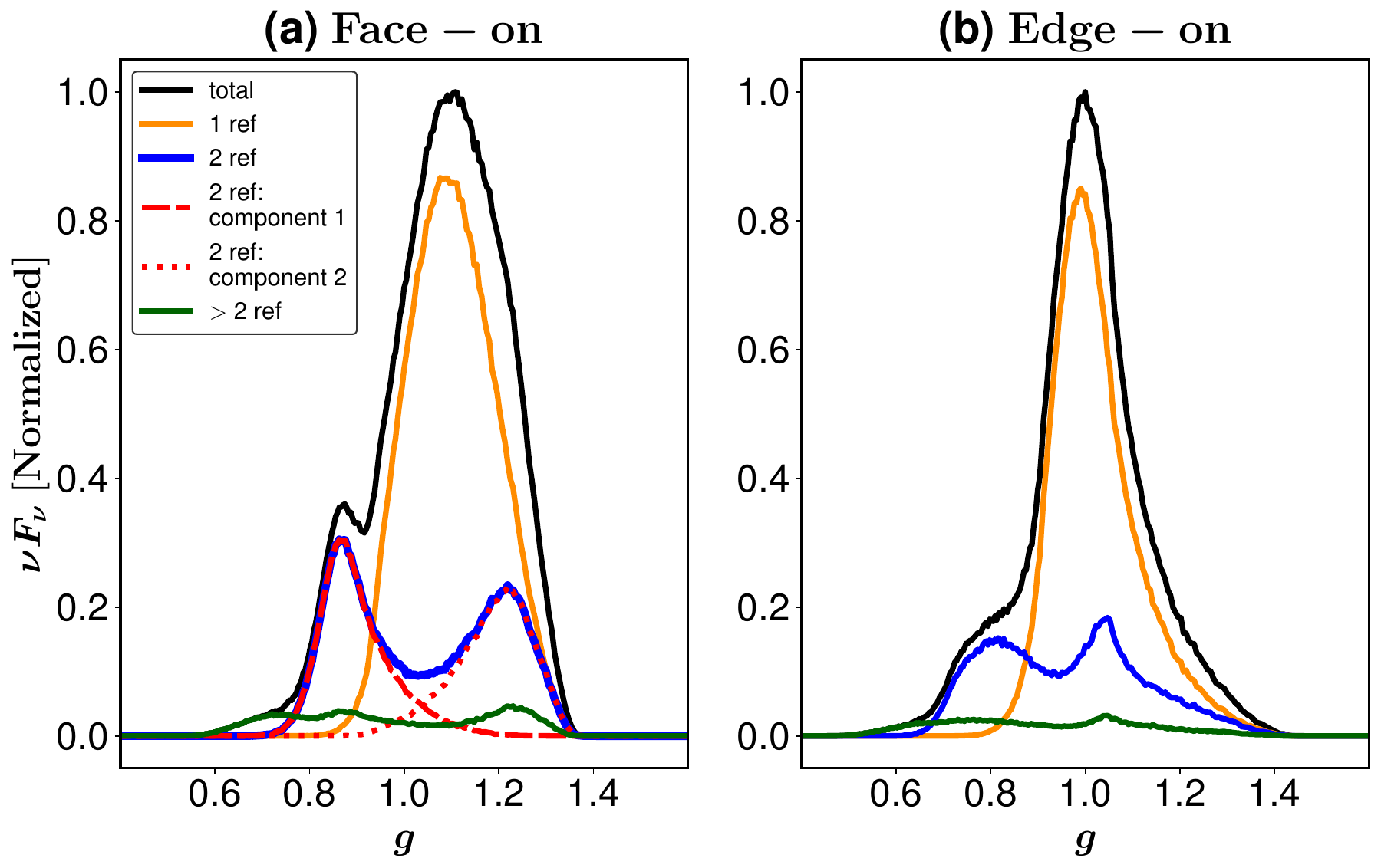}
    \caption{Fe K$\alpha$ line spectrum dependence on the observer inclination $i$. The color scheme and line styles are the same as in Fig.~\ref{FeLine}.  Panel (a) uses the default parameter set (the same as in Fig.~\ref{FeLine}), while panel (b) uses the same parameters but is viewed along the funnel wall (edge-on, $i = \theta_{\rm half} = \pi/5$).  The primary peak of the edge-on case is less blueshifted.}
    \label{incl}
\end{figure}

We take the fiducial model and compare the Fe line spectra viewed from face-on ($i=0$) and edge-on (i.e., along the funnel wall) ($i=\pi/5$) orientations in Fig.~\ref{incl}. Although $i$ varies in a limited range, the dependence on inclination is still visible. First, the most prominent difference between the two Fe line spectra is that the primary peak is less blue-shifted in the edge-on orientation compared to the face-on orientation. Additionally, one can see that in the edge-on orientation, the secondary peak is also slightly more redshifted.  
Such differences are produced because in the edge-on orientation the far side of the funnel is more exposed to the observer than the near side in the face-on orientation, while in the face-on orientation, the observer has an equal probability of viewing all sides of the funnel due to symmetry. Therefore, in the edge-on case, the averaged wind radial velocity projected along the line of sight is lower, which brings both peaks to lower energies. Overall, one can find that the energy difference between the two peaks shrinks as the observer inclination increases. For simplicity, we stick to the face-on orientation for most calculations in this work.

\subsection{The parameter space with prominent double-peak Fe line feature}\label{sec: 2D}

\begin{figure*}[!htb]
    \centering
    \includegraphics[width=0.9\linewidth]{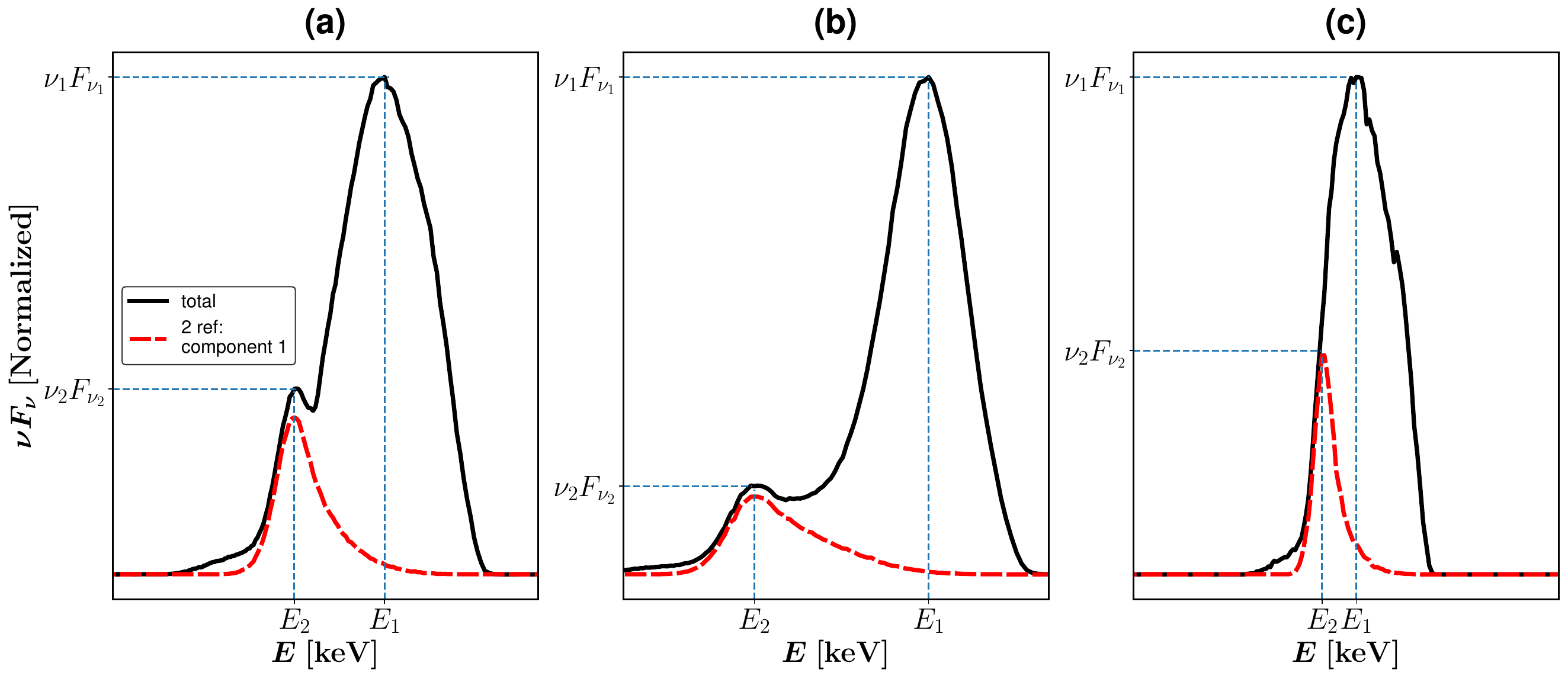}
    \caption{The illustration of the two parameters, namely, $\Delta E \equiv E_1 - E_2$ and $F_{21}\equiv \nu_2 F_{\nu_2}/\nu_1 F_{\nu_1}$, defined to quantify the significance of the double-peak line feature, based on a few characteristic modeled Fe line spectra. $E_1$ is the energy at which the total Fe line energy spectrum (black solid line) peaks and $E_2$ is the energy at which component 1 of the secondary reflection (red dashed line) peaks. $\nu_1 F_{\nu_1}$ and $\nu_2F_{\nu_2}$ are the line energy fluxes at $E_1$ and $E_2$ respectively.}
    \label{quantity}
\end{figure*}

\noindent It has been demonstrated in previous sections that the Fe K$\alpha$ lines produced by super-Eddington accretion flows can exhibit a significant double-peak feature due to multiple reflections of the photons in the funnel. Such double-peak line feature is also commonly seen in thin disk Fe K$\alpha$ lines but due to a completely different physical origin. For thin disks, the double-peak line feature is typically seen from large observer inclinations which enhances the Doppler effect induced by disk rotation. Here we conduct a systematic study to understand under which condition the double-peak Fe line feature stands out in the the super-Eddington accretion scenario.

We define two parameters to quantify the significance of the double-peak feature. The first parameter is the energy separation between the two peaks, $\Delta E \equiv E_1 - E_2$, where $E_1$ is the energy at which the whole Fe line spectrum reaches the maximum, and $E_2$ is the energy at which the redshifted component (component 1) of the secondary reflection spectrum reaches the maximum. The second parameter is the flux ratio between the secondary peak and the primary peak, $F_{21}\equiv \nu_2 F_{\nu_2}/\nu_1 F_{\nu_1}$, where $\nu_1F_{\nu_1}$ and $\nu_2F_{\nu_2}$ are the line energy fluxes measured at $E_1$ and $E_2$ respectively. These two parameters are illustrated in Fig.~\ref{quantity} for a few characteristic spectra. It can be seen that the double peak will more likely stand out when $\Delta E$ is larger or $F_{21}$ is larger (i.e., closer to 1). Therefore, we propose to use the product of $\Delta E$ and $F_{21}$ as a criterion of the significance of the double peak line feature. 

We then carry out a series of simulations to investigate how $\Delta E$, $F_{21}$ and $\Delta E\times F_{21}$ depend on the four model physical parameters $v_\infty$, $R_{\rm acc}$, $H_{\rm LP}$ and $\theta_{\rm half}$. There are $N_{v_\infty} \times N_{R_{\rm acc}} \times N_{H_{\rm LP}} \times N_{\theta_{\rm half}} = 11\times10\times11\times10 = 12100$ simulations in total. The four parameters vary within the ranges listed in Table~\ref{parameters}, each having an even distribution. The results are plotted in Fig.\,\ref{1Ds}, with Fig.~\ref{1Ds}\,(a),\,(b),\,(c) showing how $\Delta E$, $F_{21}$, and $\Delta E\times F_{21}$ depend on the four parameters respectively. In every panel, the color in each grid indicates the 2D averaged value of the quantity for the combination of the two corresponding parameters in the x and y-axis (while the other two parameters can take any value with equal chances within the designated range).  The bar plot shows the 1D averaged quantity for a fixed parameter (while the other three parameters can take any value with equal chances). 

The trends of $\Delta E$ and $F_{21}$ in Fig.\,\ref{1Ds}\,(a) and Fig.\,\ref{1Ds}\,(b) are consistent with the results discussed in Section~\ref{sec:parameters}.
$\Delta E$ and $F_{21}$ have monotonic yet opposite dependence on each of the four model parameters.  
A larger $\Delta E$ means that the wind likely has a higher terminal velocity or a faster acceleration, the funnel has a larger open angle, or the corona is placed at a higher location. However, these conditions also produce greater photon number or energy loss during the reflection process and therefore will lead to a smaller $F_{21}$. Therefore, one can expect that very extreme parameter values may not lead to prominent double peak line features, which is consistent with the results of $\Delta E \times F_{21}$ seen in Fig.~\ref{1Ds}\,(c).

We further multiply $\Delta E\times F_{21}$ by the actual double-peak feature occurrence rate (DPOC). This occurrence rate is judged based on an algorithm to recognize the local maximum of the modeled spectra, using which we can check whether the double peak feature is produced.
For a specific spectrum, its DPOC is set to be 1 as long as a secondary maximum is produced and $F_{21}$ reaches a threshold of 5\% (to account for the noise and spectral resolution in observations) or 0 otherwise. The results are shown in Fig.~\ref{1Ds}\,(d), where one can see that the 1D DPOC-weighted-$\Delta E\times F_{21}$ reaches maximum when the four parameters take intermediate values within our parameter range. Furthermore, we calculate the parameter values at which the DPOC-weighted-$\Delta E\times F_{21}$ reaches the maximum and list them in Table~\ref{table: prob}. We also give a range in favor of producing the double-peak feature within which the DPOC-weighted-$\Delta E\times F_{21}$ drops by less than $e^{-1/2}$ from the maximum value.

In conclusion, if a prominent double-peak line feature is observed from a super-Eddington accretion disk, it is most likely that the wind has a rather high terminal velocity with a relatively fast acceleration near the base. Furthermore, the lamppost corona height and funnel open angle likely take moderately large values.

\begin{deluxetable}{ccc}
\tablecaption{Parameter values when the DPOC-weighted-$\Delta E\times F_{21}$ reaches maximum or drops by less than $e^{-1/2}$ from the maximum}
\label{table: prob}
\tablehead{\colhead{parameter} & \colhead{maximum} & \colhead{favored range}} 
\startdata
$v_{\infty}$        & $0.45c$   & $[0.25c,>0.5c]$\\
$R_{\rm acc}$       & $4\,R_g$   & $[<2\,R_g ,16.8\,R_g]$\\
$H_{\rm LP}$        & $27.3\,R_g$    & $[13.4\,R_g,\ >30\,R_g]$\\
$\theta_{\rm half}$ & $40.4^\circ$  & $[25.8^\circ,\ >45^\circ]$\\
\enddata
\tablenotetext{}{``<'' and ``>'' mean that the values are outside our model parameter range}
\end{deluxetable}

\begin{figure*}[!htbp]
    \gridline{\fig{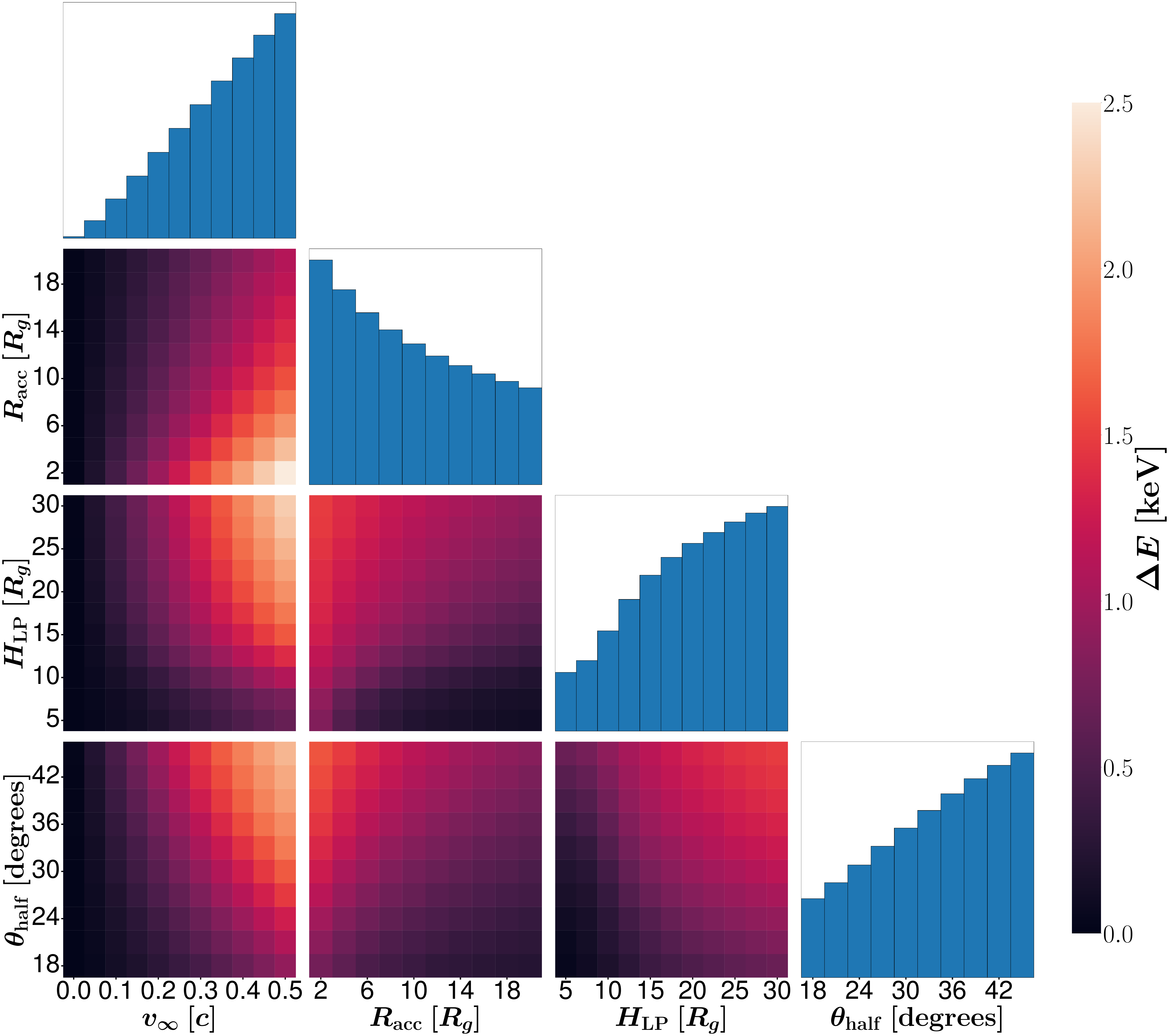}{0.65\textwidth}{(a)}}
    \gridline{\fig{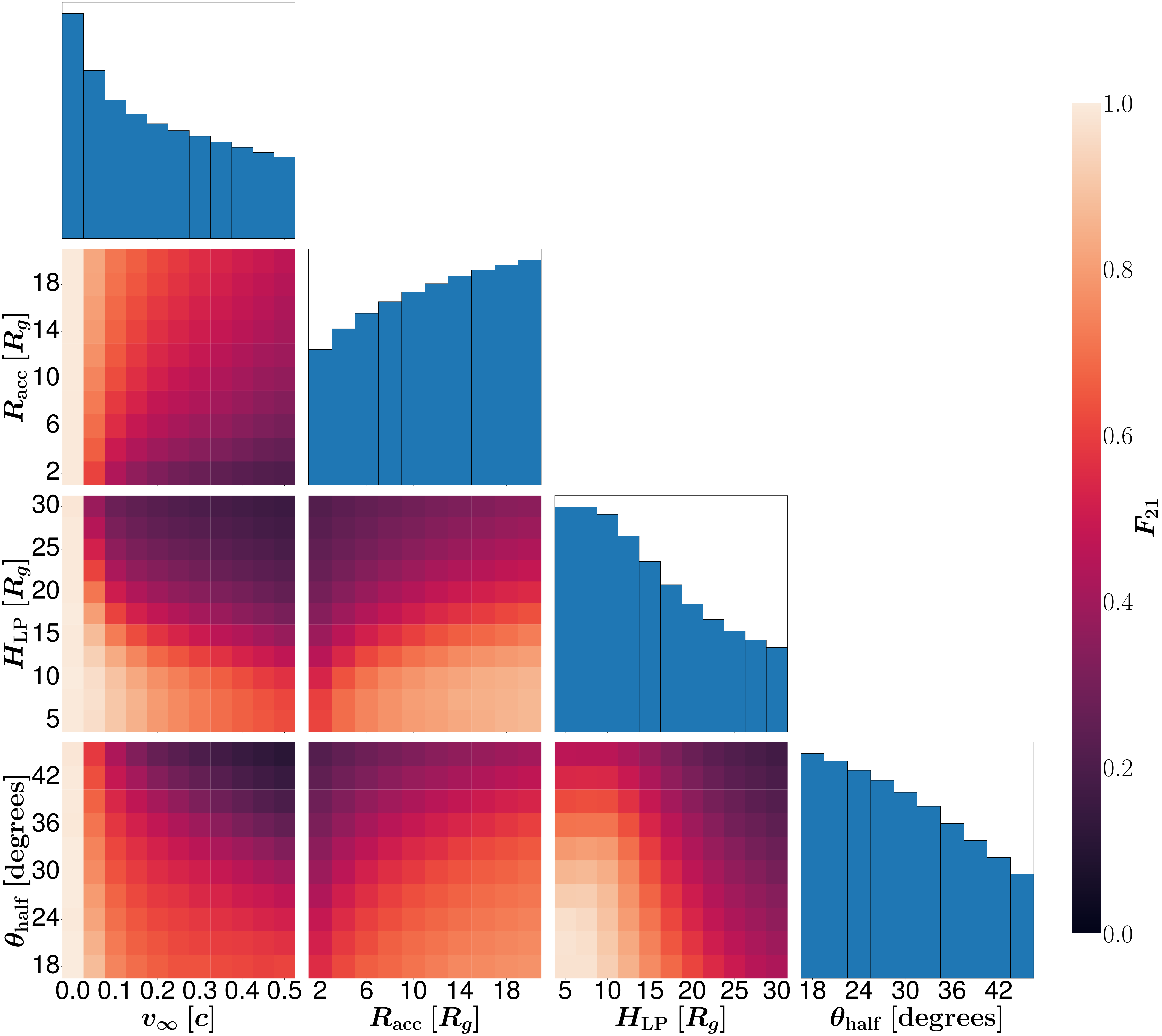}{0.65\textwidth}{(b)}}
    \vspace{-0.5cm}
    \caption{The dependence of (a) $\Delta E$, (b)$F_{21}$, (c) $\Delta E\times F_{21}$ and (d)$\Delta E\times F_{21}$ weighted by the occurrence rate of the double-peak line feature on the four model parameters $v_\infty, \ R_{\rm acc},\ H_{\rm LP}$ and $\theta_{\rm half}$. The color in each grid is the averaged 2D value of the quantity for the combination of the two parameters in the x- and y-axes (while the other two parameters take values with equal chances). The bar plots show how the averaged 1D values for the quantity depend on the parameter in the x-axis (while the other three parameters are not fixed).}
    \label{1Ds a}
\end{figure*}
\addtocounter{figure}{-1}
\begin{figure*}[!htbp]
    \gridline{\fig{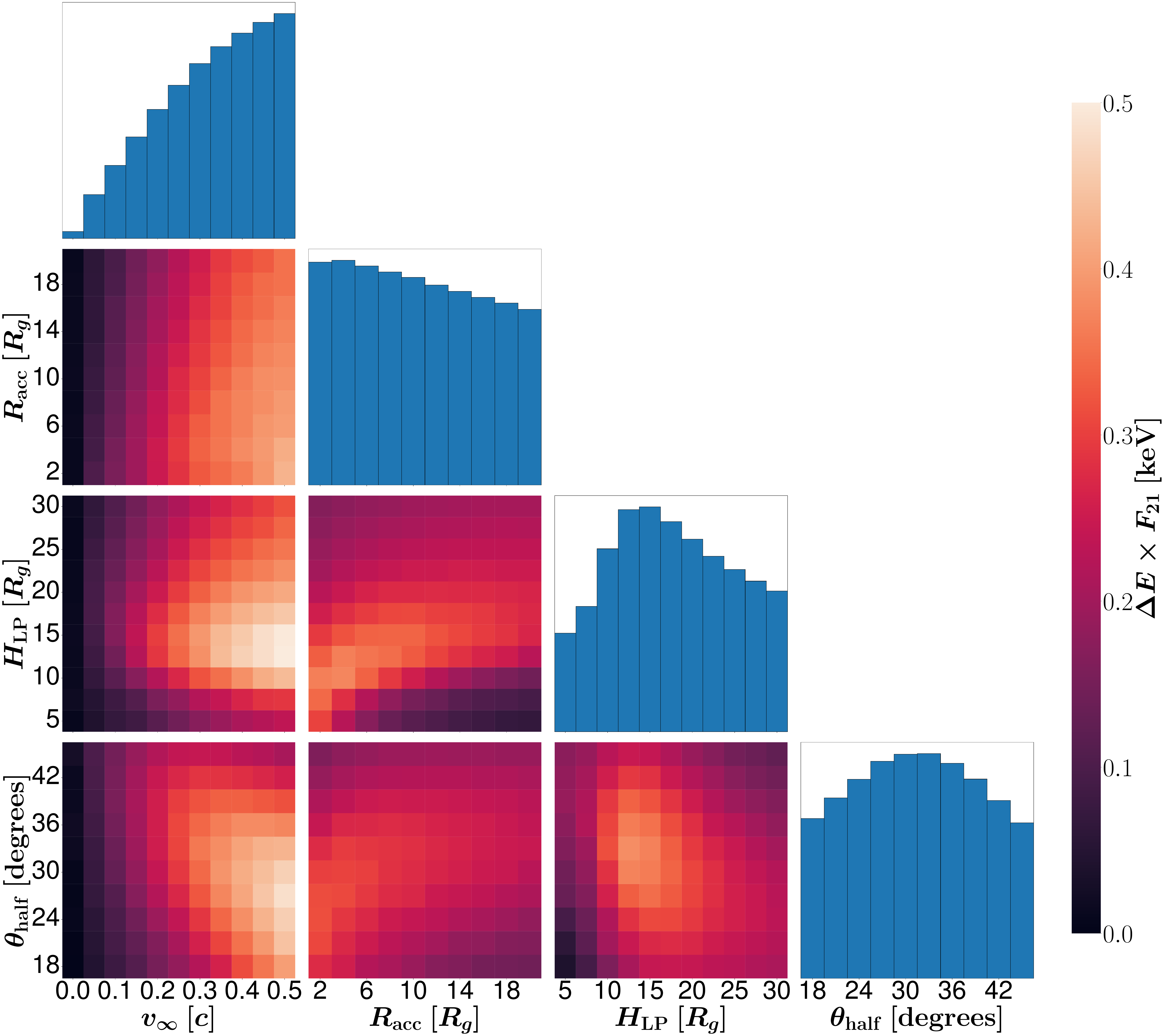}{0.65\textwidth}{(c)}}
    \gridline{\fig{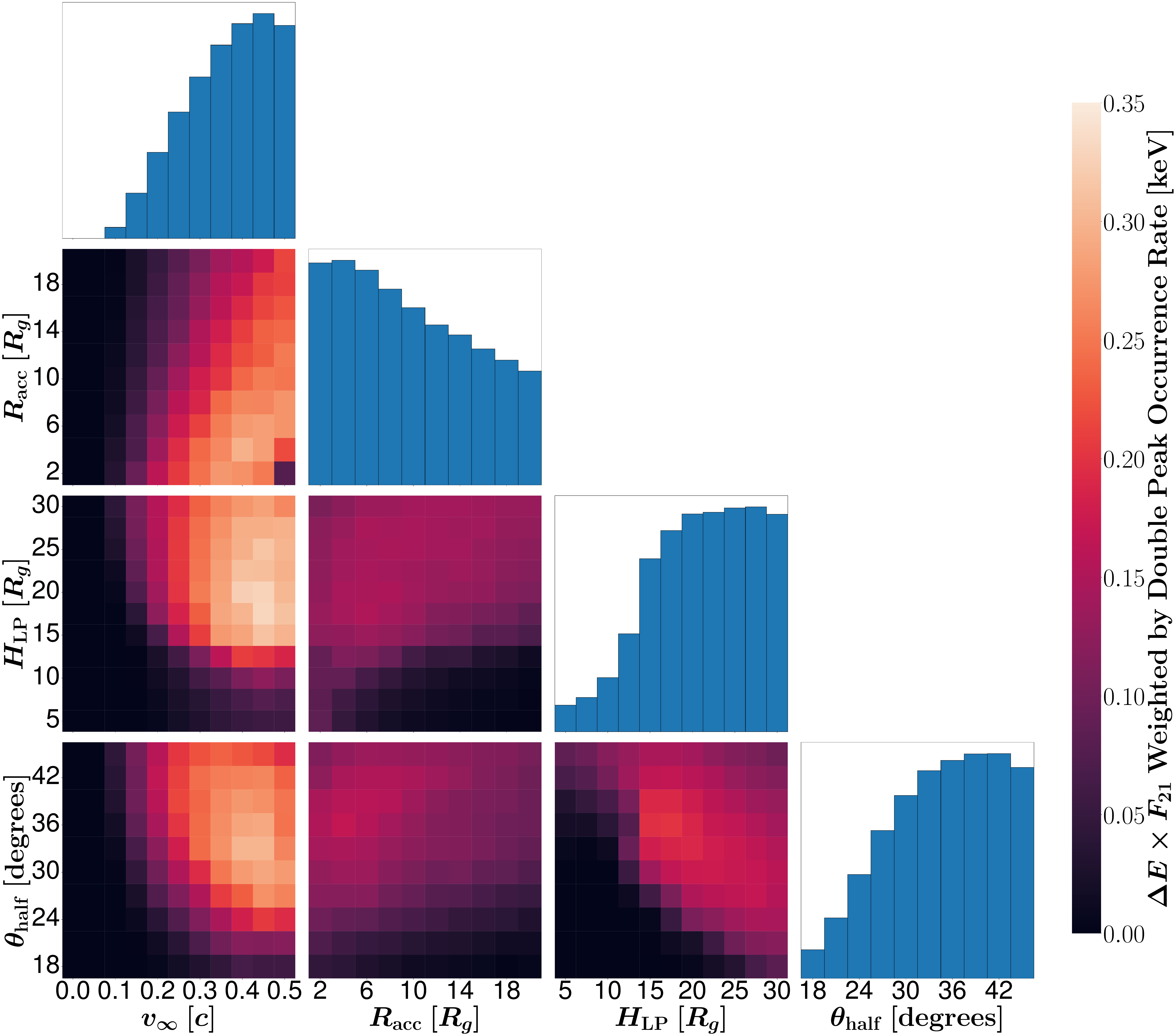}{0.65\textwidth}{(d)}}
    \vspace{-0.5cm}
    \caption{The dependence of (a)\,$\Delta E$, (b)\,$F_{21}$, (c)\,$\Delta E\times F_{21}$ and (d)\,$\Delta E\times F_{21}$ weighted by the occurrence rate of the double-peak line feature on the four model parameters $v_\infty, \ R_{\rm acc},\ H_{\rm LP}$ and $\theta_{\rm half}$. The color in each grid is the averaged 2D value of the quantity for the combination of the two parameters in the x- and y-axes (while the other two parameters take values with equal chances). The bar plots show how the averaged 1D values for the quantity depend on the parameter in the x-axis (while the other three parameters are not fixed).}
    \label{1Ds}
\end{figure*}

\begin{figure*}[htb!] 
    \centering  
	\subfigure[]{
		\includegraphics[width=0.45\linewidth]{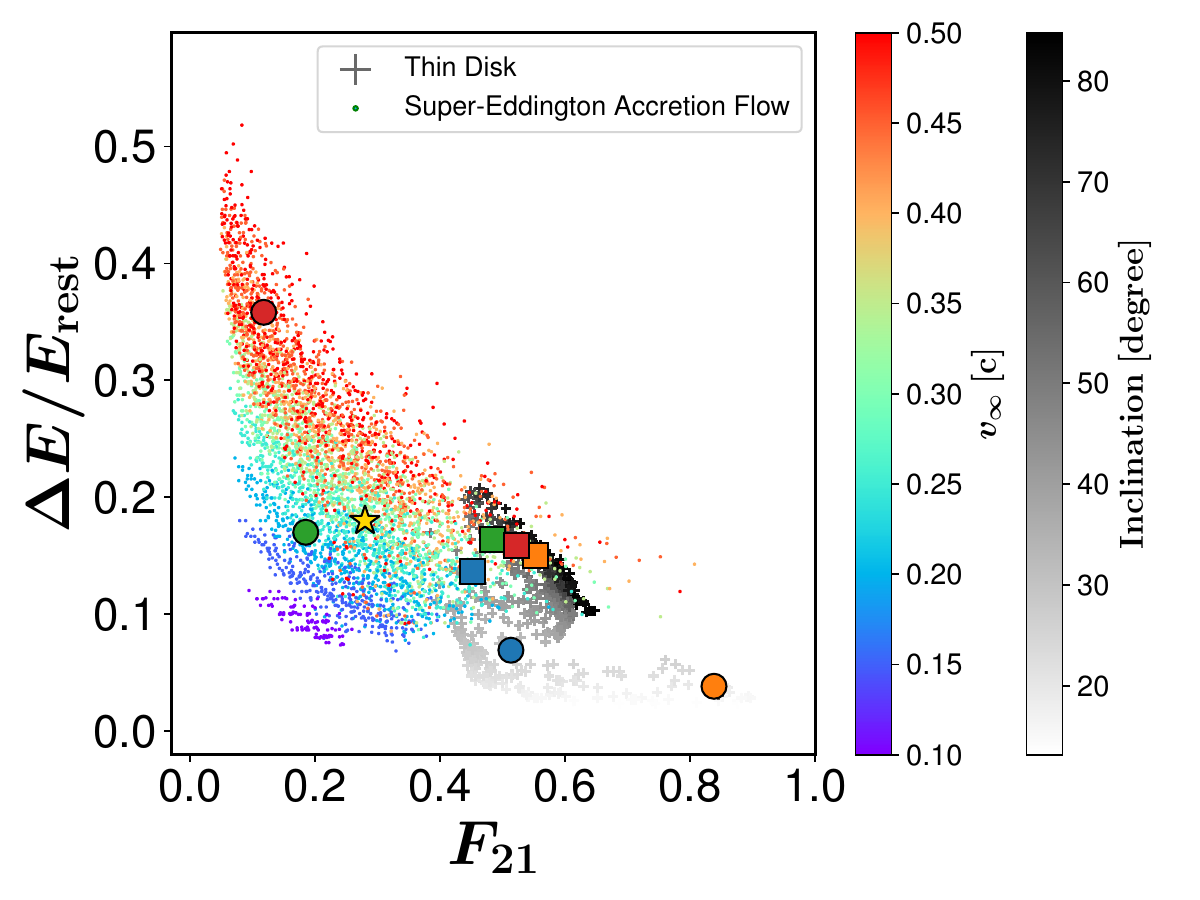}}
	\subfigure[]{
		\includegraphics[width=0.45\linewidth]{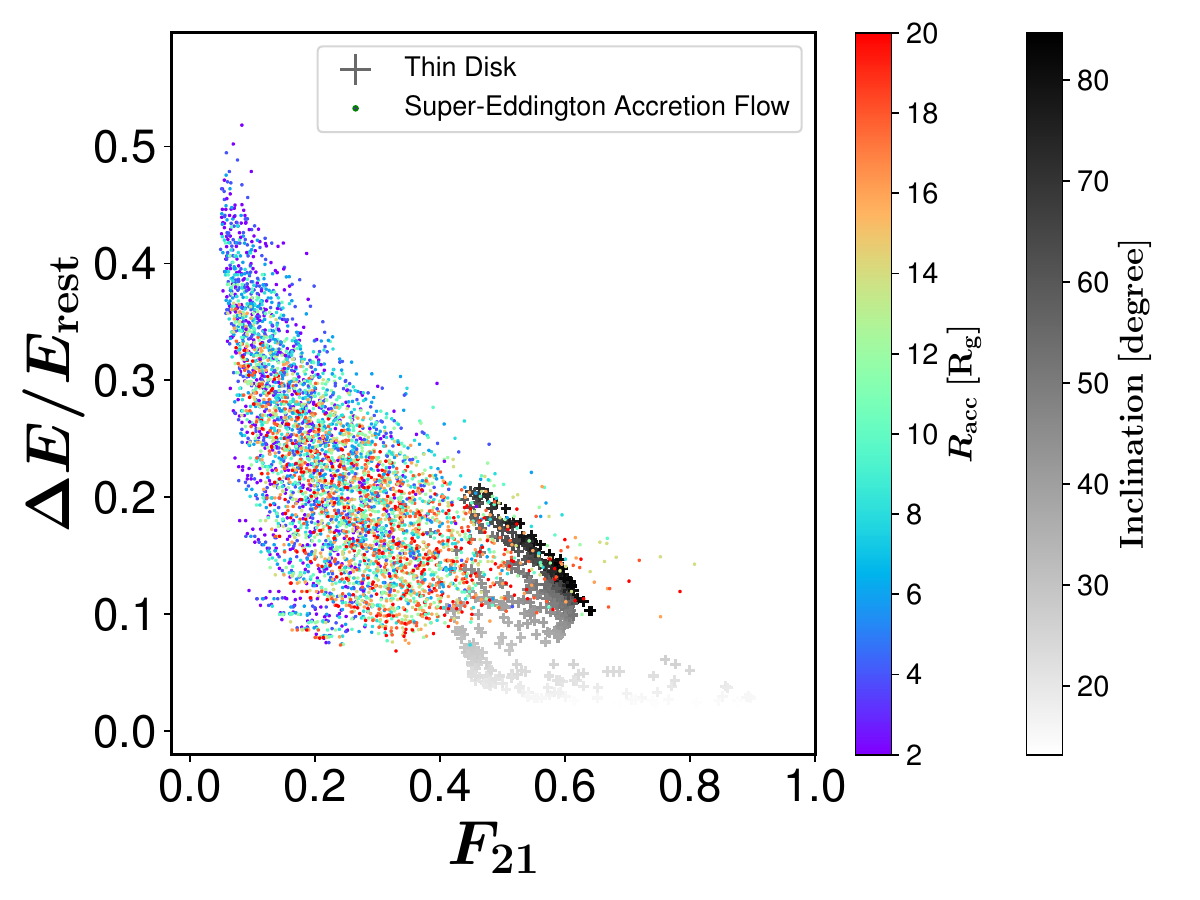}}
	\subfigure[]{
		\includegraphics[width=0.45\linewidth]{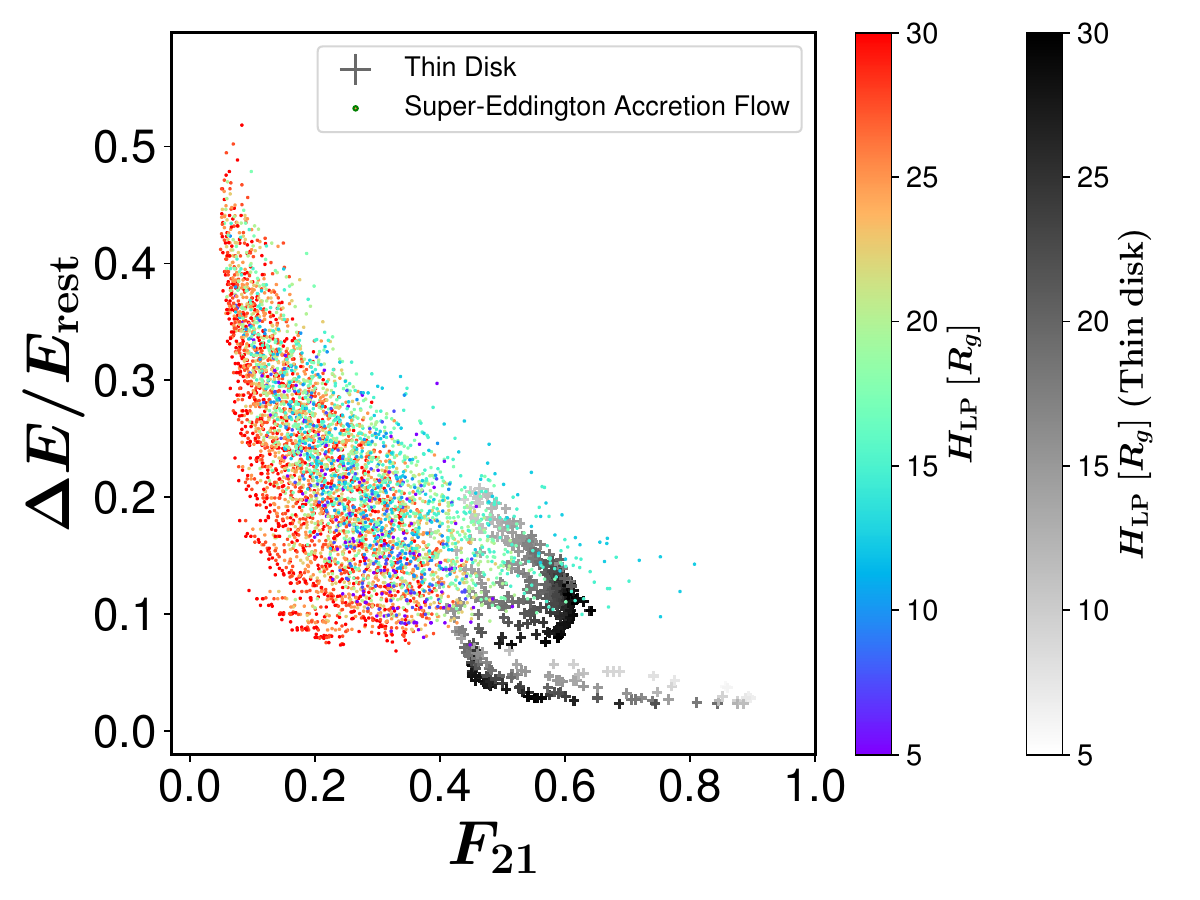}}
	\subfigure[]{
		\includegraphics[width=0.45\linewidth]{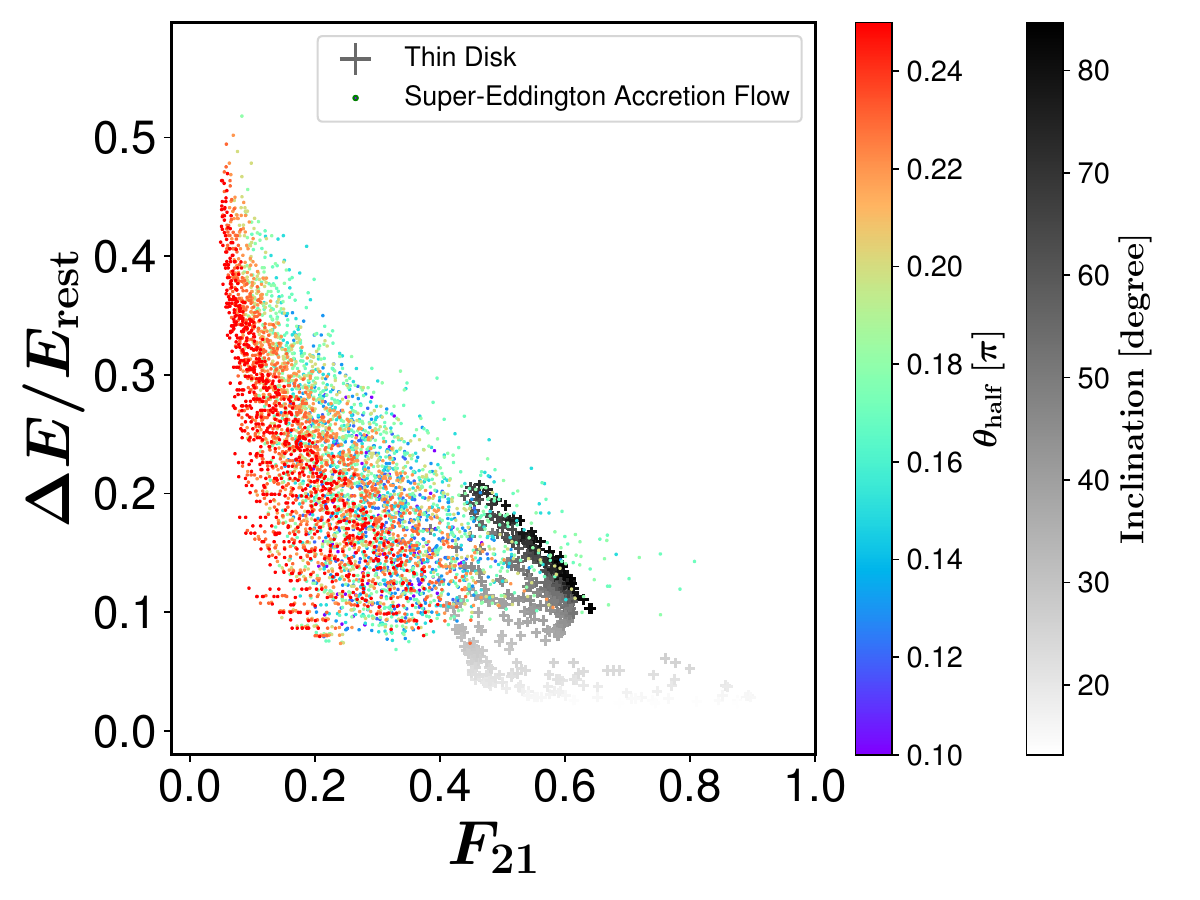}}
	\caption{The morphological comparison of the double-peak Fe lines from thin disks and super-Eddington disks. For each panel, the x-axis is the flux ratio between the secondary and primary peaks, and the y-axis is the energy difference between the two peaks normalized by the rest-frame energy of the Fe line. The super-Eddington disk Fe lines are denoted using colored points with dot marker, with the color scale representing the values of (a)\,$v_\infty$, (b)\,$R_{\rm acc}$, (c)\,$H_{\rm LP}$ and (d)\,$\theta_{\rm half}$ in each panel respectively. The thin disk Fe lines are denoted using grey points with a plus marker, with the color scale representing the inclination of the observer in panel (a), (b) and (d). In panel (c), we use the color to represent the corona height.  In panel (a) we mark the positions of the 4 lines plotted in Fig.~\ref{overlap}~(a) using circle markers and the 4 lines plotted in Fig.~\ref{overlap}~(b) using square markers. The colors of these markers are the same as the line colors in Fig.~\ref{overlap}. Furthermore, we mark the double-peaked Fe line observed in BHB 4U 1543-47 using the yellow star marker in panel (a).}
	\label{compare}
\end{figure*}

\subsection{Double-peak Fe lines from thin disks vs. super-Eddington disks}\label{morpho}
\noindent 
As the double-peak line feature can be produced from both thin-disk and super-Eddington accretion disks, it is worth exploring whether the lines from these two types of systems have any morphological differences. Again, we focus on the two parameters introduced in Section~\ref{sec:parameters}, $\Delta E$ and $F_{21}$, namely, the energy separation and flux ratio between the two peaks. Moreover, while thin disks typically produce Fe K$\alpha$ lines at rest-frame energy $E_{\rm rest} = 6.4$\,keV, and we assume that Fe K$\alpha$ lines from super-Eddington disks have $E_{\rm rest} = 6.7\,{\rm keV}$ in this work, it is possible that Fe lines at different energy levels  (e.g., 6.4, 6.7, 6.97\,keV) can be produced in either case depending on the ionization level of the gas. Therefore, we check a more generic parameter $\Delta E/E_{\rm rest}$ instead of $\Delta E$ for this morphological comparison.  

For super-Eddington disks, we stick to our sample based on the 12100 simulations as described in Section~\ref{sec:parameters}. For thin disks, we generate a sample of 400 Fe lines with the double-peak feature using the {\tt relline\_{lp}} package, which adopts a model based on a razor-thin disk with a lamppost corona \citep{Dauser13}. The corona height is randomly chosen between 5 and 30\,$R_g$ (to be consistent with the range explored for the super-Eddington disks), and the inclination of the observer is randomly chosen between $5^\circ$ and $85^\circ$. The spin of the BH is fixed to 0. Furthermore, the other parameters are set by default (e.g., the outer edge of the disk $R_{\rm out}=400\,R_g$).

We select the cases producing a double-peak Fe line feature from both samples and show the comparison of their $\Delta E/E_{\rm rest}$ and $F_{21}$ in Fig.~\ref{compare}. First and foremost, one can see that the super-Eddington and thin disk double-peak Fe lines occupy rather different $\Delta E/E_{\rm rest}$ and $F_{21}$ phase-space. Typically, the super-Eddington disk Fe lines have larger energy separation but a lower flux ratio between the two peaks as compared to the thin disk counterparts. This means that it is possible to tell the disk accretion state and geometry based on the analysis of the double-peak Fe line itself. Second, one can see that for super-Eddington disks the dependence of $\Delta E/E_{\rm rest}$ and $F_{21}$ on the four different parameters $v_\infty$, $R_{\rm acc}$, $H_{\rm LP}$ and $\theta_{\rm half}$ is consistent with the results in Section~\ref{sec:parameters}. The dependence on $v_\infty$ is the strongest, which indicates that there is a promising prospect of constraining the super-Eddington wind velocity through the Fe K$\alpha$ line profiles. Third, for thin disks  $\Delta E/E_{\rm rest}$ and $F_{21}$ also depend on $H_{\rm LP}$, as shown in Fig.~\ref{compare} (c). However, no monotonic trends can be seen. Fourth, the dependence of $\Delta E$ on the observer inclination $i$ is opposite in the two cases. For thin disks, as $i$ increases, $\Delta E$  generally increases. For super-Eddington disks, as demonstrated in Section~\ref{sec:inclination}, $\Delta E$ tends to decrease with increasing $i$, although the dependence should be mild due to the limited variation of $i$ within the small funnel open angle. Last but not least, the two samples have an overlap in the parameter space where $\Delta E/E_{\rm rest}$ is on the smaller side for the super-Eddington disk sample but on the larger side for the thin disk sample.

\begin{figure*}[!htb]
    \centering
    \includegraphics[width=0.7\linewidth]{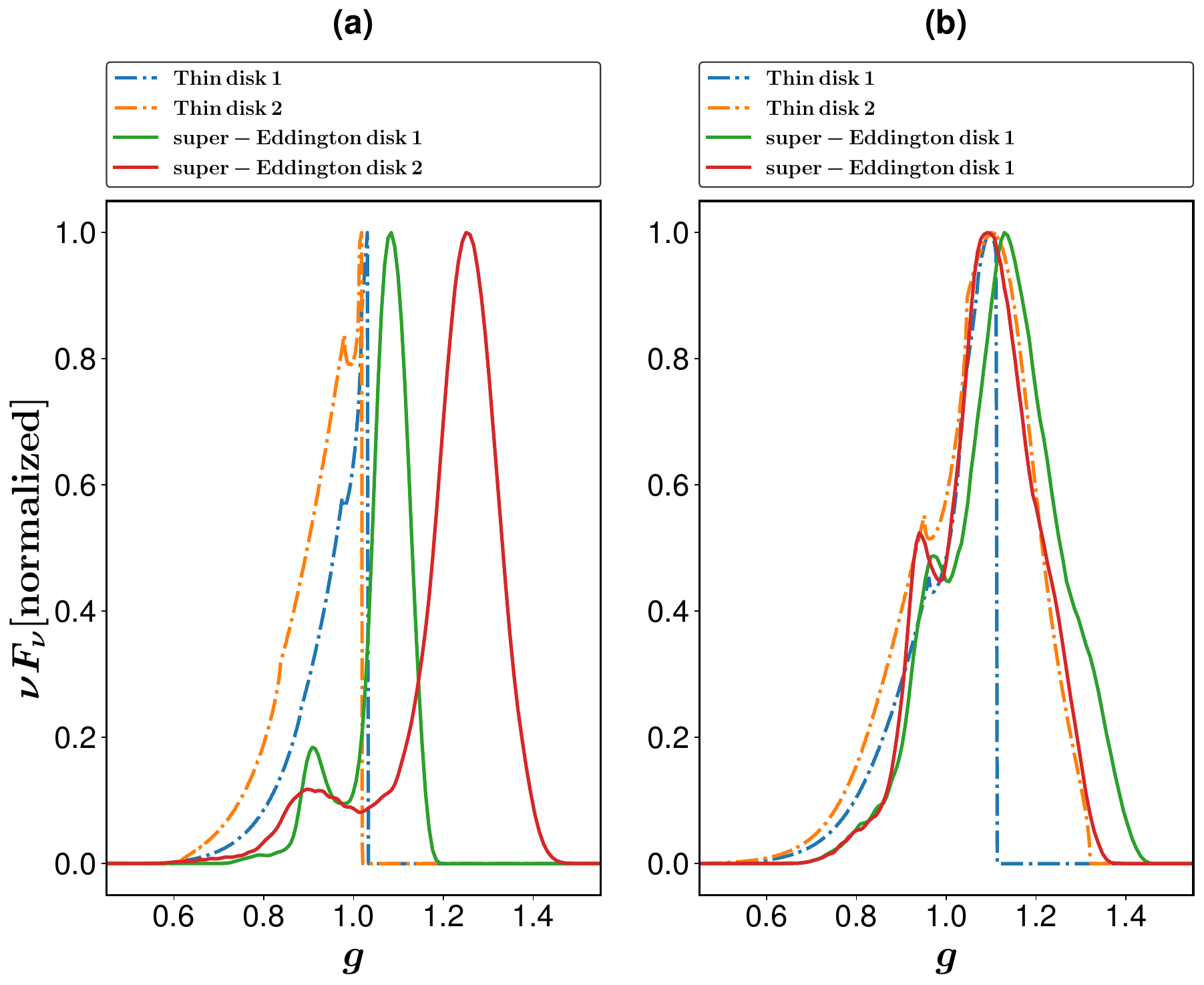}
    \caption{Comparison of the Fe K$\alpha$ line profiles with similar $\Delta E/E_{\rm rest}$ and $F_{21}$ from the thin-disk sample (dash-dotted line) and super-Eddington simulations (solid-line). We show the line profiles that have very different $\Delta E/E_{\rm rest}$ and $F_{21}$ in panel (a) and those have similar $\Delta E/E_{\rm rest}$ and $F_{21}$ in panel (b). The model parameters are listed below. (a) thin-disk 1: $H_{\rm LP}=11.3\,R_g,\, i=25.0^{\circ}$; thin disk 2: $H_{\rm LP}=6.8\,R_g,\, i=19.3^{\circ}$; super-Eddington disk 1: $v_\infty=0.2c$, $\theta_{\rm half}=0.22\pi$, $R_{\rm acc}=8\, R_g$ and $H_{\rm LP} = 25\, R_g$; super-Eddington disk 2: $v_\infty=0.4c$, $\theta_{\rm half}=0.18\pi$, $R_{\rm acc}=4\, R_g$ and $H_{\rm LP} = 25\, R_g$. (b) thin-disk 1: $H_{\rm LP}=14.2\,R_g,\, i=44.9^{\circ}$; thin disk 2: $H_{\rm LP}=18.6\,R_g,\, i=70.3^{\circ}$; super-Eddington disk 1: $v_\infty=0.35c$, $\theta_{\rm half}=0.15\pi$, $R_{\rm acc}=12\, R_g$ and $H_{\rm LP} = 17.5\, R_g$; super-Eddington disk 2: $v_\infty=0.35c$, $\theta_{\rm half}=0.17\pi$, $R_{\rm acc}=14\, R_g$ and $H_{\rm LP} = 17.5\, R_g$.}
    \label{overlap}
\end{figure*}

We also plot a few characteristic Fe K$\alpha$ line profiles from both types of disks in Fig.~\ref{overlap} for a more direct comparison. In Fig.~\ref{overlap}~(a) we randomly select 2 lines from super-Eddington disks and 2 lines from thin disks which occupy very different phase spaces in Fig.~\ref{compare}. One can see that the super-Eddington Fe lines can generally have larger blueshifts and larger energy separation between the two peaks. Then we further randomly select 2 lines from super-Eddington disks and 2 lines from thin disks that occupy the overlapping phase space region in Fig.~\ref{compare} and plot their line profiles in Fig.~\ref{overlap}~(b). Here one can see that although these lines have similar $\Delta E/E_{\rm rest}$ and $F_{21}$, there can still be some differences in the exact line profiles. The primary peak of the Fe line from a thin disk can have a skewed shape when the observation inclination is low, while that from a super-Eddington disk always has a relatively symmetric shape. However, there are indeed cases when the thin disk also produces less skewed Fe line profiles when the observer inclination is high, and under such scenarios, more careful analysis will be needed to distinguish the lines from the two types of disks. This also highlights the importance of conducting further modeling to study the whole reflection spectrum from super-Eddington accretion flows beyond the Fe K$\alpha$ lines. 

\section{Fitting the F\lowercase{e} K$\alpha$ line observed in BHB 4U 1543--47}\label{sec:fit}
\noindent 
The Galactic BHB 4U1543--47 was first discovered by \citet{Matilsky72} and experienced a few outbursts in 1983, 1992, 2002, and 2021 \citep{Kitamoto82,Harmon92,Park04,Buxton04,Negoro21a}. For the year 2021 burst \citet{Negoro21b} reported a $2\,{-}\,10$ keV flux of $1.96\times 10^{-7} \rm{erg\,cm^{-2}\, s^{-1}}$ and a corresponding bolometric luminosity of $1.34\times 10^{39} \rm{erg\, s^{-1}}$, which is comparable to the Eddington luminosity of the black hole with a mass of 9.4~$M_\odot$ at a distance of 7.5 kpc. \citet{Jin24} further reported that the peak luminosity captured by Insight-HXMT reached $\sim2.3\,L_{\rm Edd}$.

\citet{Jin24} and \citet{Zhao24} analyzed the spectra of the 2021 burst and showed that the spectra at and after the peak of the burst consist of a soft disk blackbody component and a hard component from the corona which can be modeled by a simple power-law or a thermally comptonized continuum. On top of that, they found that the excess emissions at $\sim 5-7\, \rm{keV}$ with an absorption feature at $\sim 7-12\, \rm{keV}$ can be best explained as a characteristic Fe K$\alpha$ reflection signature. Interestingly,  both Insight-HXMT and \nustar observations showed that a double-peak feature appeared in the Fe K$\alpha$ line profile during this state. Therefore, the observed Fe K$\alpha$ line is likely produced when 4U1543--47 is in a super-Eddington accretion state, which motivates us to fit this Fe line using our model as a proof of concept.

\begin{figure}[!htbp]
    \centering
    \includegraphics[width=1.0\linewidth]{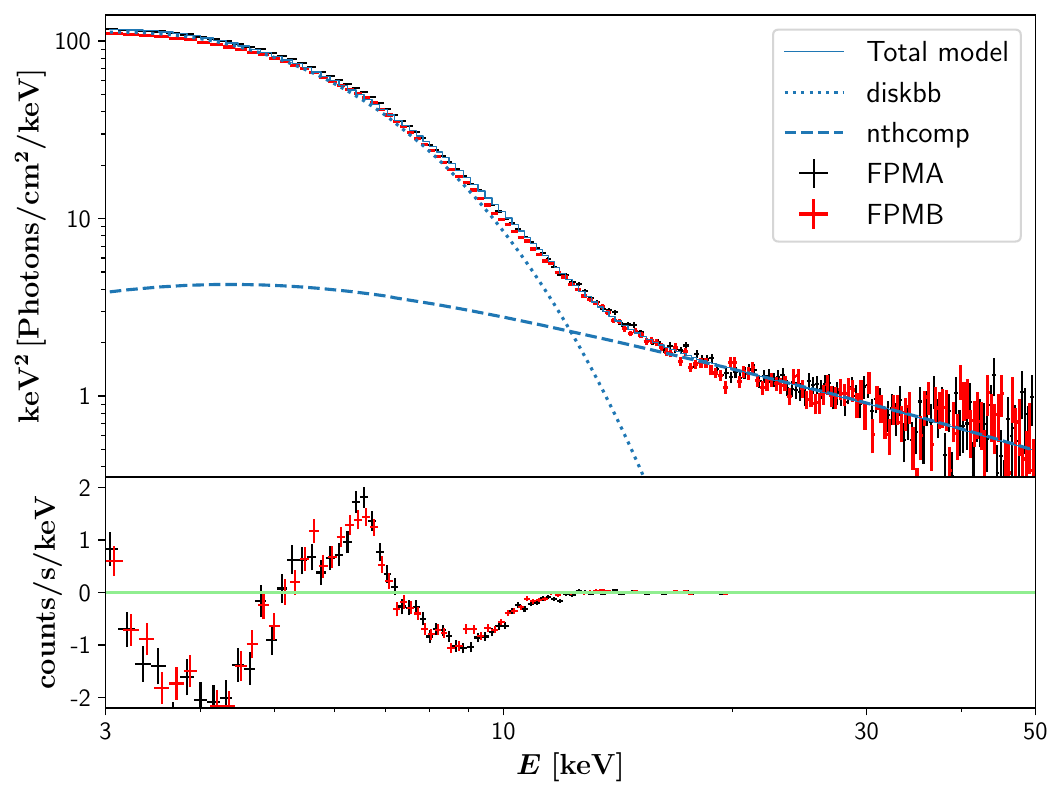}
    \caption{Spectra from a \nustar observation of 4U1543--47 during the 2021 ourburst phase fitted using the \texttt{tbabs*(diskbb+nthcomp)} model. In the top panel, black and red data points show the unfolded spectra from FPMA and FPMB. For a better visual effect, we only show the modelled FPMA spectrum (blue solid curve) with its decomposed components \texttt{diskbb} (blue dotted line) and \texttt{nthcomp} (blue dashed line. The bottom panel shows the residual between the observed and the modelled spectra.}
    \label{contfit_diskbb}
\end{figure}

\begin{figure}[!htbp]
    \centering
    \includegraphics[width=0.75\linewidth]{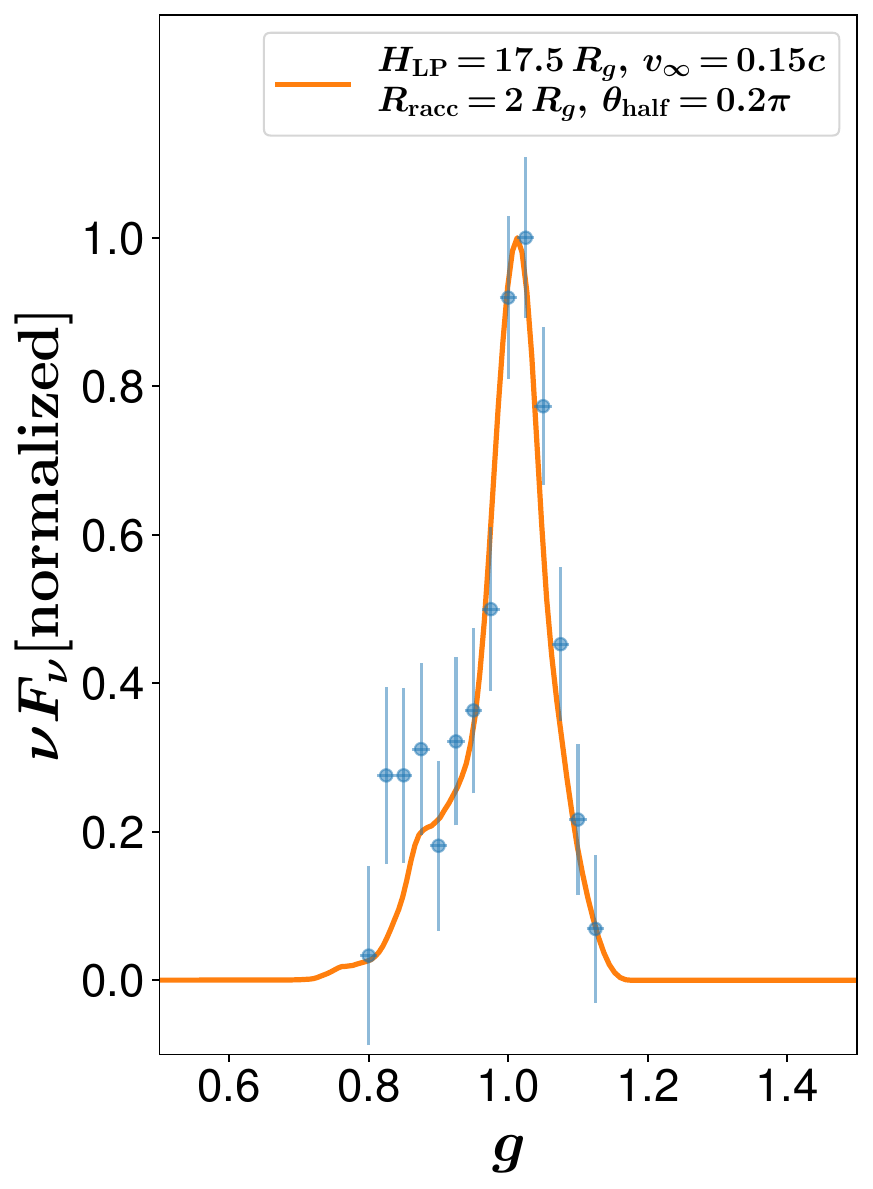}
    \caption{Best-fit model for the Fe K$\alpha$ line which corresponds to the positive residual in the $5-8$ keV region in the lower panel of Fig. \ref{contfit_diskbb}. The blue data points are the normalized \nustar residual in the line region. The orange curve is the best-fit super-Eddington Fe K$\alpha$  model with multiple reflection. The fitted model parameters are listed in the legend.}
    \label{fit_diskbb}
\end{figure}

We use the \nustar data from one epoch when the source is in the super-Eddington state (obsID: 80702317002) as shown in Fig.~\ref{contfit_diskbb}. We first fit the observed X-ray spectrum within the range of $3-79$ keV with a \texttt{tbabs*(diskbb+nthcomp)} model by ignoring the energy range $\left[3.4, 12.0\right]$ keV to exclude the Fe line and absorption feature. The best fit parameters are shown in Table~\ref{tab:contfitdiskbb}. Then we obtain the difference between the observed spectrum and the fitted continuum spectrum, and use the positive residual data to represent the Fe line. The residual spectrum from the FPMA instrument (converted to $g$ with the assumption that this Fe K$\alpha$ line has a rest frame energy of 6.4 keV) is shown in Fig. \ref{fit_diskbb}, which exhibits a clear double peak feature. We further estimate the $\Delta E/E_{\rm rest}$ and $F_{21}$ of this observed Fe K$\alpha$ line and show its phase-space position in Fig.~\ref{compare}\,(a) (denoted using yellow star). One can see that the small $F_{21}$ value supports that this source likely falls into the super-Eddington disk regime. Lastly, we fit this residual to our super-Eddington Fe K$\alpha$ line model. The best-fit model is shown in Fig.~\ref{fit_diskbb}, which has $v_\infty = 0.15c$, $\theta_{\rm half}=0.2\pi$, $R_{\rm acc}=2\, R_g$ and $H_{\rm LP} = 17.5\, R_g$. These parameters, especially the wide funnel open angle, are reasonable for a mildly super-Eddington accretion flow.

\begin{deluxetable*}{c c c}

	\caption{Best-fit parameters for the \nustar observation using the \texttt{tbabs*(diskbb+nthcomp)} model} \label{tab:contfitdiskbb}
	
    \tablehead{\colhead{Model Component (XSPEC Model)} & \colhead{Parameter (Units)} & \colhead{Value}}

	\startdata
	Galactic absorption (\texttt{tbabs}) & $N_{H,\,\mathrm{gal}}$ ($10^{22}$ cm$^{-2}$) & 0.46\tablenotemark{a} \\
	Multi-black body (\texttt{diskbb}) & $T_{\rm in}$ (keV) & $1.202\pm 0.007$ \\
	& ${\rm norm}$ & $8577.9\pm 175.5$ \\
	Thermally comptonized continuum (\texttt{nthcomp}) & 
	$kT_e$ & $1000\tablenotemark{a}$\\
	& $kT_{bb}$ & $=T_{\rm in}$\\
    & inp type & 1 \\
    & $z$ & 0\tablenotemark{a} \\
    & ${\rm norm}$ & $1.7\pm 0.2$ \\
    Cross-Calibration (\texttt{const}) &$C_\mathrm{FPMA}$ & 1\tablenotemark{a} \\
	& $C_\mathrm{FPMB}$ & $0.941\pm0.002$\\
	Fit Statistic & $\chi^2_\nu/{\rm d.o.f.}$& 328.94/302\\
    \enddata
    \tablenotetext{a}{Parameter was fixed when fitting.}
\end{deluxetable*}

Alternatively, we fit the observed \nustar spectrum with a different model \texttt{tbabs*(bbody+powerlaw+gaussian+gaussian)*smedge}. A blackbody model has been previously used to approximate the emissions from super-Eddington accretion disks in literature \citep{Masterson22, Yao24arXiv}. Here the two Gaussian functions are used to approximate the double peak feature in the Fe K$\alpha$ line energy range. The best fit parameters are shown in Table~\ref{tab:contfit}. We next obtain the difference between the observed spectrum and the \texttt{tbabs*(bbody+powerlaw)*smedge} components and assume this residual represents the Fe K$\alpha$ line. The residual spectrum is shown in  Fig.~\ref{fit}, which also exhibits a clear double peak feature and $\Delta E/E_{\rm rest}$ and $F_{21}$ of this observed line are similar to the previous results. The residual of FPMA spectrum is fitted again to our super-Eddington Fe K$\alpha$ line model. The best-fit model is shown in Fig.~\ref{fit}, which gives $v_\infty = 0.35c$, $\theta_{\rm half}=0.18\pi$, $R_{\rm acc}=14\, R_g$ and $H_{\rm LP} = 17.5\, R_g$. The similarity of the fitted parameters from both models demonstrates the robustness of our fitting procedure.

\begin{figure}[!htbp]
    \centering
    \includegraphics[width=1.0\linewidth]{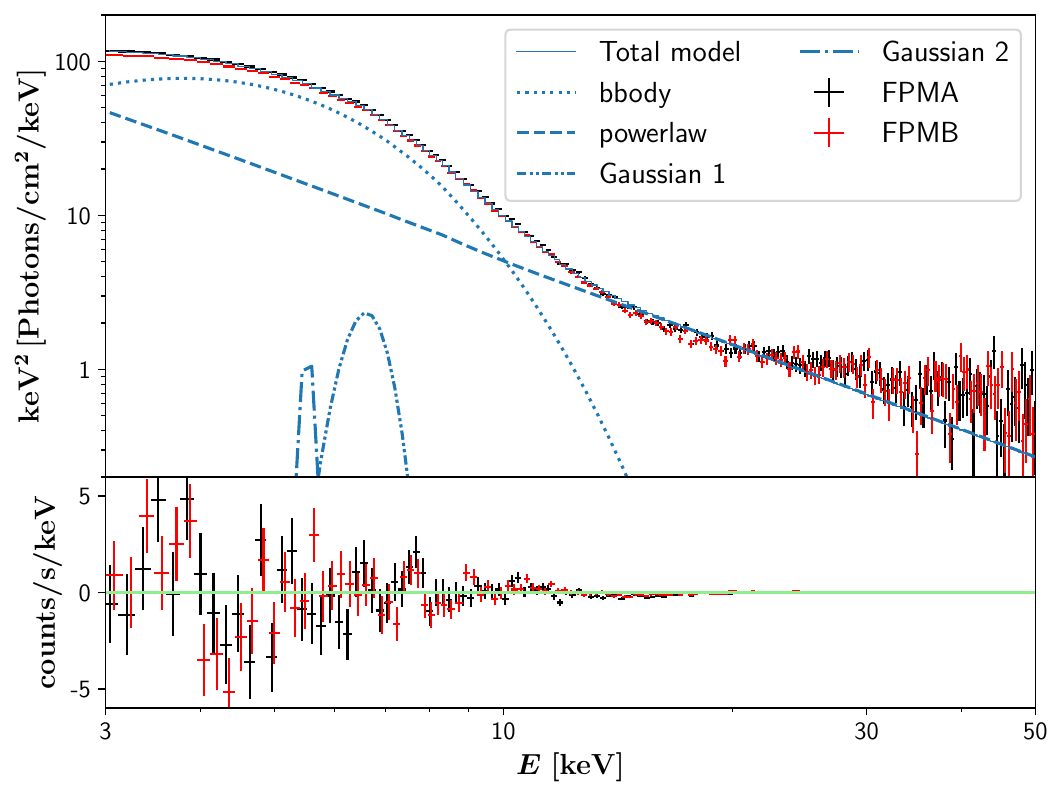}
    \caption{The same spectra as in Fig. \ref{contfit}  fitted but using an alternative model \texttt{tbabs*(bbody+powerlaw+gaussian+gaussian)*smedge}. In the top panel, besides the observed data we show the total modelled FPMA spectrum  (blue solid curve) and its decomposed components \texttt{bbody}(blue dotted curve), \texttt{powerlaw} (blue dashed curve), and \texttt{gaussian} (blue dash-dotted curves). The bottom panel shows the residual between the observed and the modeled spectra.}
    \label{contfit}
\end{figure}
\begin{figure}[!htbp]
    \centering
    \includegraphics[width=0.75\linewidth]{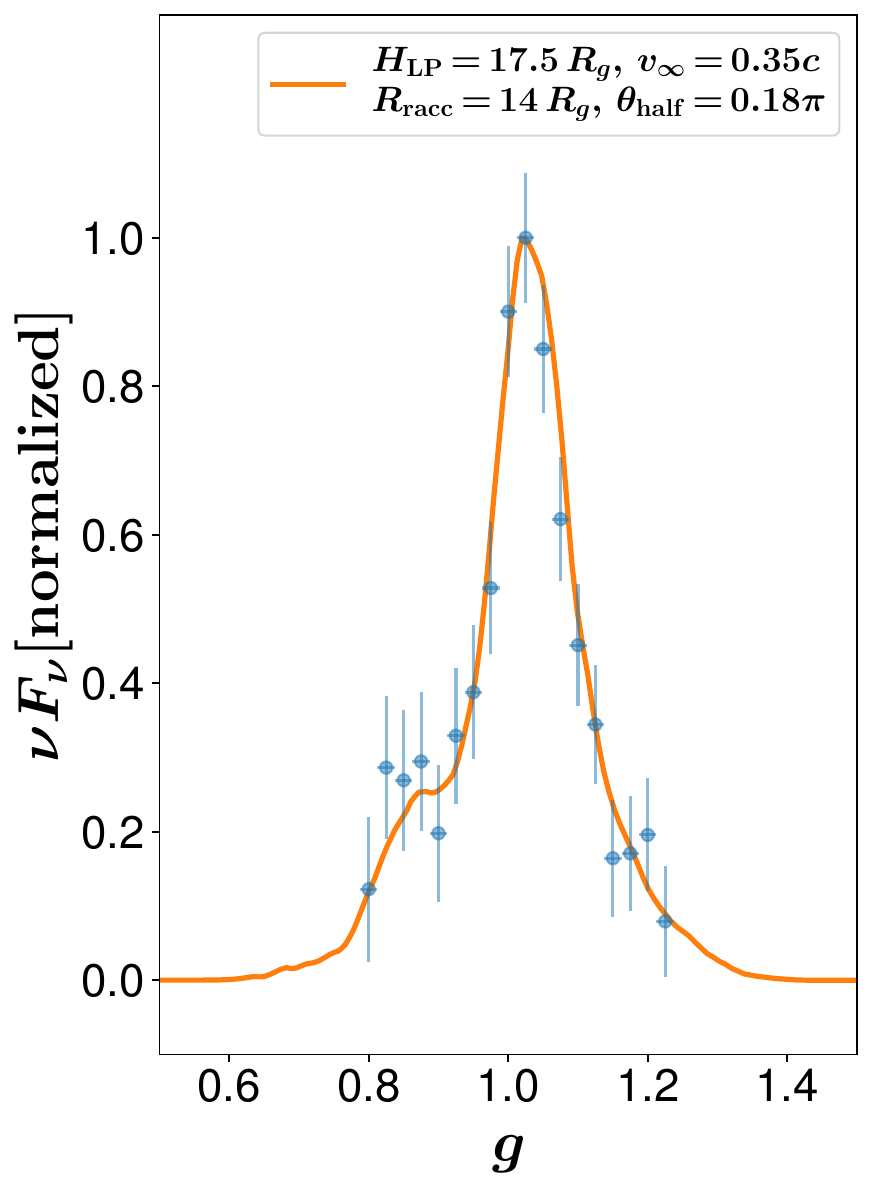}
    \caption{
     Best-fit model for the Fe K$\alpha$ line which corresponds to the positive residual in the $5-8$ keV region in the lower panel of Fig. \ref{contfit} after taking out the two \texttt{gaussian} components. The line styles and colors are the same as in Fig. \ref{fit_diskbb}.}
    \label{fit}
\end{figure}

\begin{deluxetable*}{c c c}

	\caption{Best-fit parameters for the \nustar observation using the alternative model} \label{tab:contfit}
	
    \tablehead{\colhead{Model Component (XSPEC Model)} & \colhead{Parameter (Units)} & \colhead{Value}}

	\startdata
	Galactic absorption (\texttt{tbabs}) & $N_{H,\,\mathrm{gal}}$ ($10^{22}$ cm$^{-2}$) & 0.46\tablenotemark{a} \\
	Blackbody (\texttt{bbody}) & $kT$ (keV) & $0.9502\pm 0.0011$ \\
	& ${\rm norm}$ & $2.066\pm 0.007$ \\
	Power Law (\texttt{powerlaw}) & $\Gamma$ & $3.853\pm0.006$\\
	& ${\rm norm}$ & $380.5\pm 5.3$\\
	Gaussian 1 (\texttt{gaussian}) & $E_l\, {\rm (keV)}$ & $6.55\pm0.02$ \\
    & $\sigma$ (keV)& $0.055\pm 0.005$ \\
    & ${\rm norm}$ & $0.054\pm 0.005$ \\
	Gaussian 2 (\texttt{gaussian}) & $E_l\, {\rm (keV)}$ & $5.52\pm0.05$ \\
    & $\sigma$ (keV)& $0.08\pm 0.09$ \\
    & ${\rm norm}$ & $0.011\pm 0.003$ \\
    Absorption Edge (\texttt{smedge}) & $E_c$ (keV) & $8.27\pm0.09$\\
    &$\tau_{\rm max}$ & $0.21\pm0.02$\\
    &$\alpha$& -2.67\tablenotemark{a} \\
    &smearing width (keV) & $3.5$\tablenotemark{a}\\
	Cross-Calibration (\texttt{const}) & $C_\mathrm{FPMA}$ & 1\tablenotemark{a} \\
	& $C_\mathrm{FPMB}$ & $0.9412\pm0.0007$\\
	Fit Statistic & $\chi^2_\nu/{\rm d.o.f.}$& 759.95/318\\
    \enddata
    \tablenotetext{a}{Parameter was fixed when fitting.}
\end{deluxetable*}

\section{Summary and Discussion}\label{sec: summary}
\noindent X-ray reflection from super-Eddington accretion flows around black holes are understudied. Current theoretical studies have only considered the disk geometry obtained from one specific disk simulation and single reflection of the coronal X-ray photons \citep{Thomsen19, Thomsen22a}. 
In this study, we extend the study of this topic by considering a variety of super-Eddington funnel geometries and wind kinematics and allowing the X-ray photons to scatter multiple times in the funnel. We focus on investigating the energy spectrum profiles of Fe K$\alpha$ lines produced due to the irradiation of the super-Eddington wind by the coronal photons and summarize the main results below.

\begin{itemize}

\item The most interesting result is that the Fe K$\alpha$ lines from super-Eddington accretions flows can have two peaks at optimal conditions. The primary peak is blueshifted and the secondary peak is redshifted (Fig.~\ref{FeLine}). We note that such double-peak Fe lines, when observed, are commonly interpreted to result from the rotation of geometrically thin accretion disks viewed from relatively high inclinations. However, the double-peak line features from super-Eddington accretion flows have a completely different origin -- the observer needs to have an almost face-on orientation to look into the optically thin funnel and the secondary peak is induced by multiple reflections of photons in the funnel. 

\item The Fe line profile from a super-Eddington accretion flow is sensitive to the kinematics of the wind and in particular its terminal velocity, and also depends on the wind acceleration profile, funnel open angle, and the height of the corona (Fig.~\ref{vinf}, \ref{Racc}, \ref{h}, \ref{open}).  

\item It is also worth mentioning that in certain parameter spaces, the Fe K$\alpha$ line may not exhibit a prominent double peak feature. In this scenario, the two spectral peaks can merge into one and a single broad, blueshifted line with a rather symmetric shape can be produced, or the line can have a blueshifted peak with a plateau feature in the red wing (e.g., Fig.~\ref{vinf}).

\item We have constrained the parameter space for producing a prominent double peak Fe line feature (Fig.~\ref{1Ds} and Table~\ref{table: prob}). If a double peak Fe line is observed from a super-Eddington accreting system, the wind likely has relatively quick accretion near the base and a relatively high terminal velocity, the corona is placed above $\sim 10\,R_g$, and the funnel open angle is moderately large. 

\item We have compared the profiles of the double-peak Fe lines from super-Eddington accretion flows and thin disks. We find that for the former, the energy separation between the two peaks can be larger, while the flux ratio between the secondary peak and the primary peak tends to be smaller compared to the latter (Fig.~\ref{compare}). These morphological differences in the Fe lines can be used to distinguish between the two types of accretion disks, although there is a parameter space where the double-peak Fe line profiles from both systems are similar.

\item As a proof of concept, we fit the spectrum of a bright BHB 4U 1543-47 which produced a double-peak Fe K$\alpha$ line feature during a super-Eddington outburst phase. The fitted results indicate that multiple reflection can be responsible for producing the Fe line secondary peak seen in this source, and the fitted parameters are consistent with a mildly super-Eddington accretion flow model (Fig. \ref{fit}).

\end{itemize}

In this work, we have systematically investigated how super-Eddington funnel geometry and wind kinematics shape the Fe K$\alpha$ energy spectrum, based on a simplified cone geometry for the funnel. There are a few caveats: 
1) Super-Eddington accretion flows likely have X-ray spectra are usually very soft, possibly with photon indexes $\Gamma$ larger than 3 or 4, while in this study we adopt $\Gamma=2$ for simplicity (although we also find the choice of $\Gamma$ barely affects the Fe K$\alpha$ line profile). 
2) We assume a constant density throughout the wind. In practice, the wind launched from super-Eddington accretion disks should have a density structure and different ionzation states depending on the location.
3) Following the previous point, Fe K$\alpha$ lines have complicated ionization structures. Super-Eddington accretion flows have ionization levels and therefore can produce hot iron lines at 6.7 or 6.97 keV  \citep{Ballantyne01}. In the scenario when both lines are produced, their broad line profiles can merge into a more blurred double peak line profile (see Appendix \ref{FeIonization}).4) We adopt the simple lamppost corona model for simplicity, while recent X-ray polarization studies show that the corona geometry is likely more complicated \citep[e.g.][]{Krawczynski22,Gianolli23}. Nonetheless, the calculations we have conducted here based on simplified treatments have demonstrated that X-ray reflection spectroscopy including the Fe K$\alpha$ line profile has great potentials in constraining the funnel geometry and wind kinematics in super-Eddington accretion.

We have mainly focused on the energy spectrum of the Fe K$\alpha$ line in this work. However, we have also shown that other features of the reflection spectrum, such as the Fe absorption edge, also reveal information about the super-Eddington wind and funnel and therefore are worthy of further detailed study. A complete study should include more sophisticated radiative transfer physics, such as the ionization state of the gas, the absorption and emission processes due to other elements, and the spectral shape of the corona emission, etc.
In future works we plan to carry more comprehensive radiative transfer modellings and fully explore more X-ray reflection spectroscopic features from super-Eddington accretion flows. In particular, a broad feature around 1\,keV has been observed in super-Eddington flows formed in X-ray TDEs \citep{Masterson22,Yao24arXiv} and  narrow line Seyfert 1 galaxies \citep{Xu22}, which has been suggested to be also produced by X-ray reflection. If such lines are also produced at the base of the funnel, they can also likely go through multiple reflections and produce similar line profiles as the Fe K$\alpha$ line discussed here. 

Studies like ours are highly motivated by the emerging observations of X-ray reflection and time lag signatures from super-Eddington or close-to-Eddington accretion flow \citep{Kara16b, Mundo20, Yao22, Masterson22}. Also, more super-Eddington accreting systems such as X-ray strong TDEs have been and are expected to be observed from new transient surveys by \srge \citep{eROSITA21} and the Einstein Probe telescope \citep{Yuan22}, while the high-resolution spectroscopic observations of such systems can be made using the \xmm \citep{XMM}, \nustar \citep{NuSTAR}, XRISM \citep{XRAM, XRISM}, Athena \citep{Athena}, AXIS \citep{Reynolds23,Arcodia24}, etc. More theoretical studies on not only the entire reflection energy spectrum but also the combination of the energy-and-time analysis will be very useful for understanding the properties of the accretion disk, wind, and corona production around black holes in the context of super-Eddington accretion.

\begin{acknowledgments}
\noindent We thank the anonymous referee for constructive comments. We thank C. Bambi, Y. Fang, Y. Inoue,  C. Jin, P. Kosec, T. Kwan, Y. Ma, K. Ohsuga, I. Papadakis, Z. Shi, J. Wang, Y. Xu, Y. Zeng, and J. Zhang for helpful comments and discussions. ZZ, LT, and LD acknowledge  support from the National Natural Science Foundation of China and the Hong Kong Research Grants Council (HKU12122309, HKU17305920, HKU27305119, N\_HKU782/23). TD acknowledges support by the Deutsche Forschungsgemeinschaft (project number 443220636, research unit FOR 5195). This research was supported in part by grant no. NSF PHY-2309135 to the Kavli Institute for Theoretical Physics (KITP).
\end{acknowledgments}

\bibliography{multi}{}
\bibliographystyle{aasjournal}

\appendix
\restartappendixnumbering

\section{Validity of the Assumption of Photon Elastic Scattering}\label{elastic}
\noindent We test the validity of considering only elastic scattering. If Compton scattering process is considered, the photon energy $E_\gamma$ before scattering, the photon energy $E_{\gamma^\prime}$ after scattering, and the scattering angle $\theta$ should satisfy:
\begin{equation}
    E_{\gamma^\prime}=\frac{E_\gamma}{1+(E_\gamma/m_e c^2)(1-\cos\theta)}.
\end{equation}

Here we adopt the low energy limit of Klein-Nishina formula for the differential cross section of the scattering process:
\begin{equation}
    \frac{{\rm d}\sigma}{{\rm d}\Omega} \approx \frac{1}{2}r_e^2(1+\cos^2\theta).
\end{equation}

\begin{figure}[!htbp]
    \centering
    \includegraphics[width=0.9\linewidth]{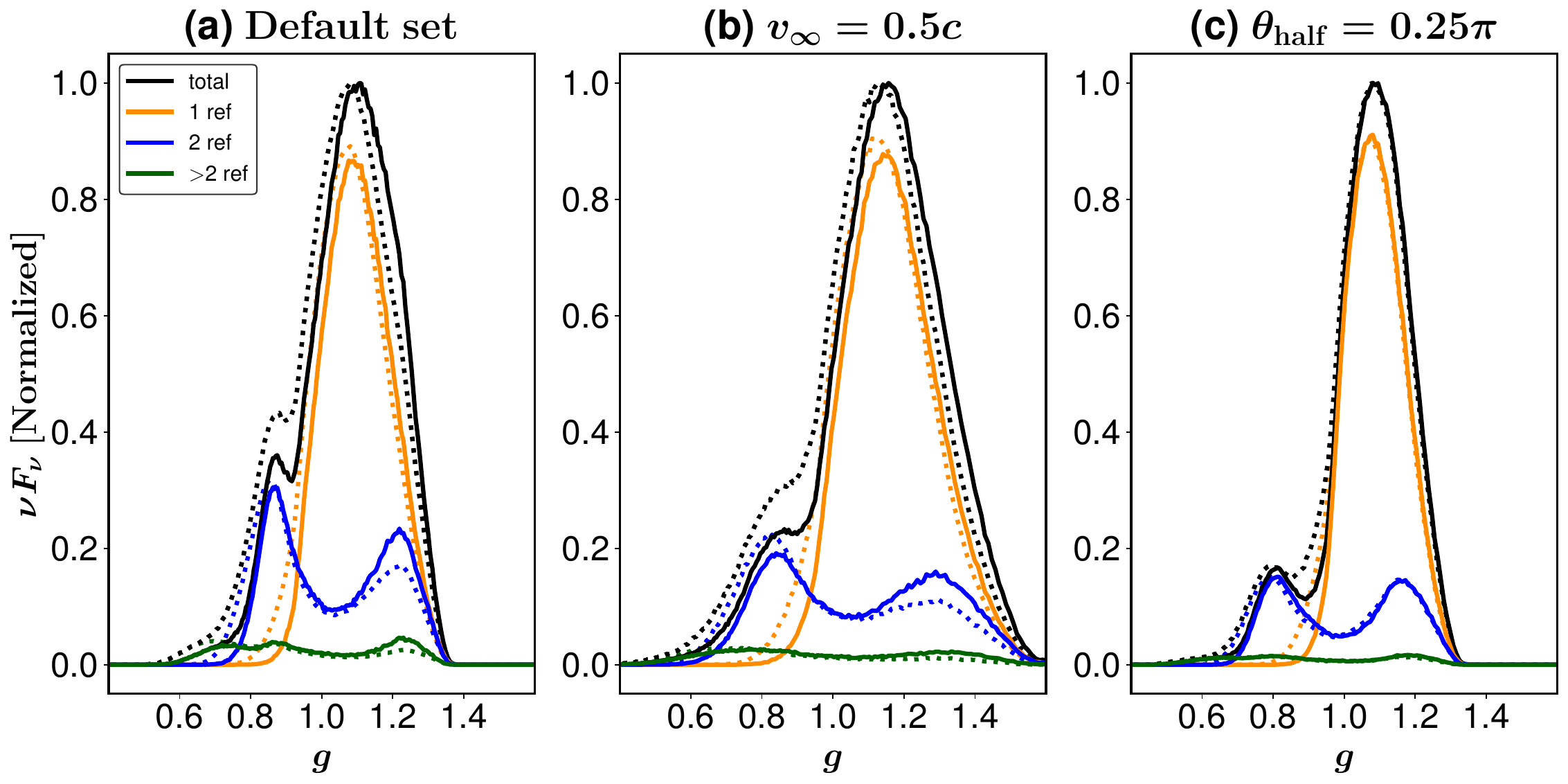}
    \caption{A comparison of the Fe line profiles calculated using elastic scattering (solid line) vs. Compton scattering (dotted line). The model parameters of (a), (b), (c) are the same as Fig.~\ref{FeLine}~(b), Fig.~\ref{vinf}~(c), Fig.~\ref{open}~(c) separately.}
    \label{compton}
\end{figure}

The results of using Compton scattering and elastic scattering, for the fiducial model and two other models, are compared in Fig.~\ref{compton}. The double-peak feature is still visible, and the first-order results remains similar. Component 1 of the secondary reflection has a slightly larger flux due to photon energy loss in Compton scattering.

\restartappendixnumbering
\section{The absorption edge}\label{absorption}
\begin{figure*}[!htbp]
    \centering
    \includegraphics[width=1.0\linewidth]{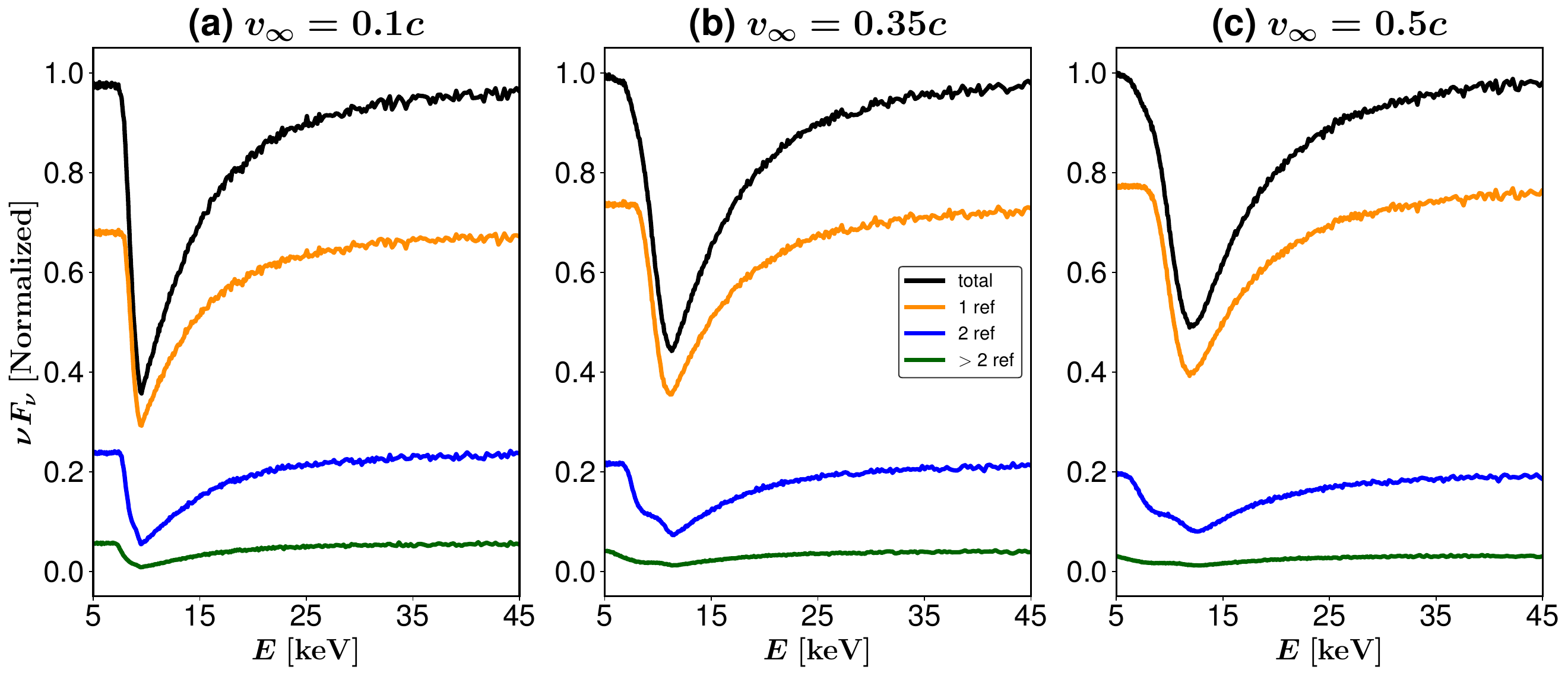}
    \caption{The dependence of the absorption edge profile on the wind terminal velocity $v_\infty$. The black line shows the spectrum of all escaped photons that have only gone through scatterings in the wind, which is decomposed into several components: orange - photons that have undergone one scattering;  blue – photons that have undergone two scatterings; green – photons that have undergone three or more scatterings. Panel (b) is the result of the default set of model parameters (the same as in Fig.~\ref{FeLine}), while panel (a) and (c) have different $v_\infty$. As $v_\infty$ increases, the absorption edge becomes less skewed, since the trough of the absorption edge is more blueshifted and the edge starts at lower energy.}
    \label{cont_vinf}
\end{figure*}

\begin{figure*}[!htbp]
    \centering
    \includegraphics[width=1.0\linewidth]{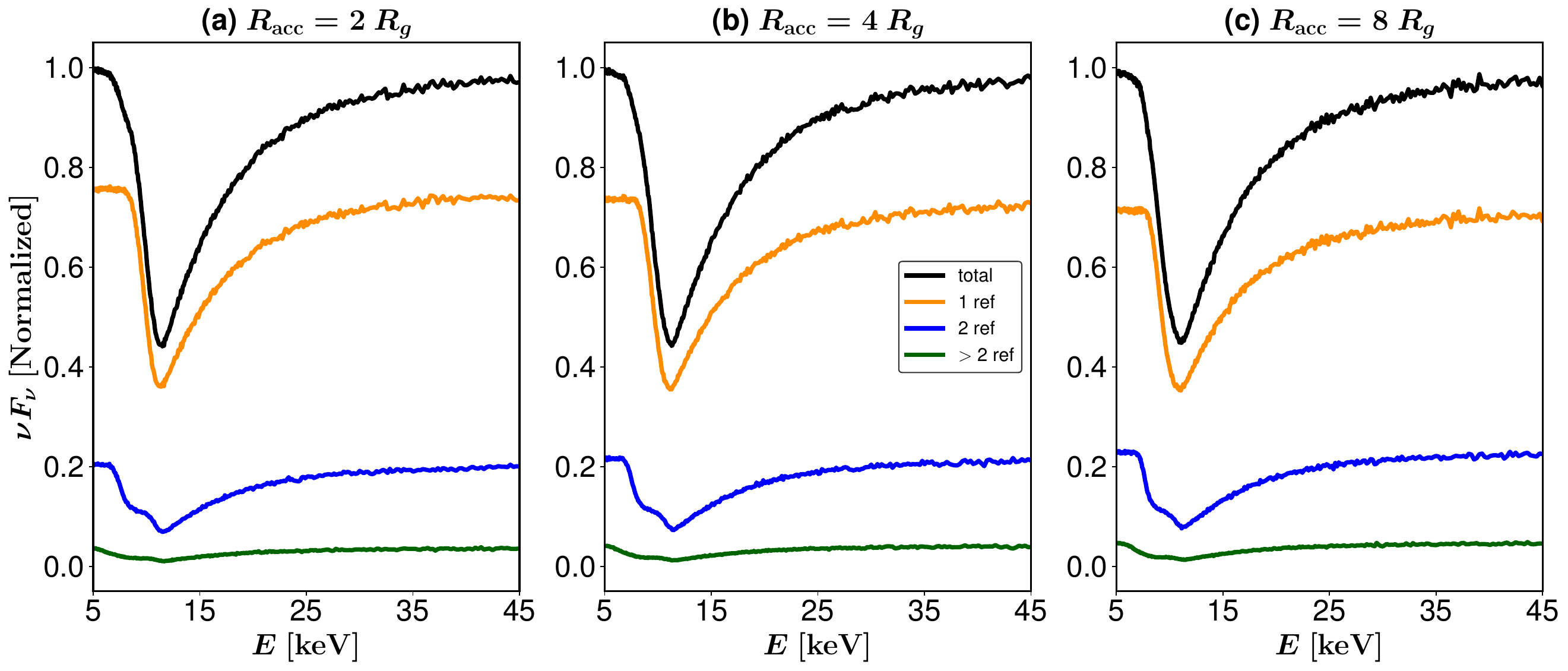}
    \caption{The dependence of the absorption edge profile on the wind acceleration radius $R_{\rm acc}$. The color scheme and line styles are the same as in Fig.~\ref{cont_vinf}. Panel (b) is the result of the default set of model parameters, while panel (a) and (c) have different $R_{\rm acc}$. As $R_{\rm acc}$ decreases, the absorption edge becomes slightly less skewed.}
    \label{cont_Racc}
\end{figure*}

\begin{figure*}[!htbp]
    \centering
    \includegraphics[width=1.0\linewidth]{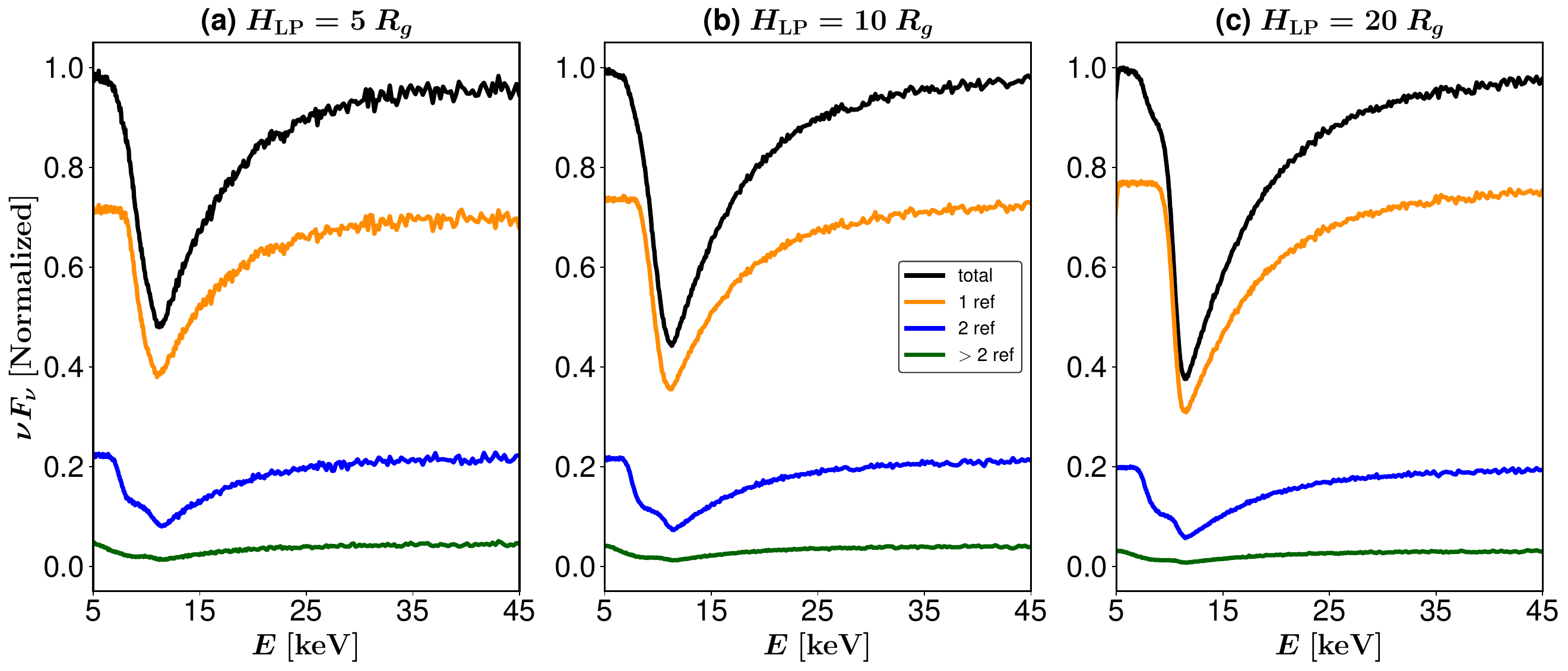}
    \caption{The dependence of the absorption edge profile on the height of corona $H_{\rm LP}$. The color scheme and line styles are the same as in Fig.~\ref{cont_vinf}. Panel (b) is the result of the default set of model parameters, while panel (a) and (c) have different $H_{\rm LP}$. As $H_{\rm LP}$ decreases, the absorption edge becomes less skewed.}
    \label{cont_hlp}
\end{figure*}

\begin{figure*}[!htbp]
    \centering
    \includegraphics[width=1.0\linewidth]{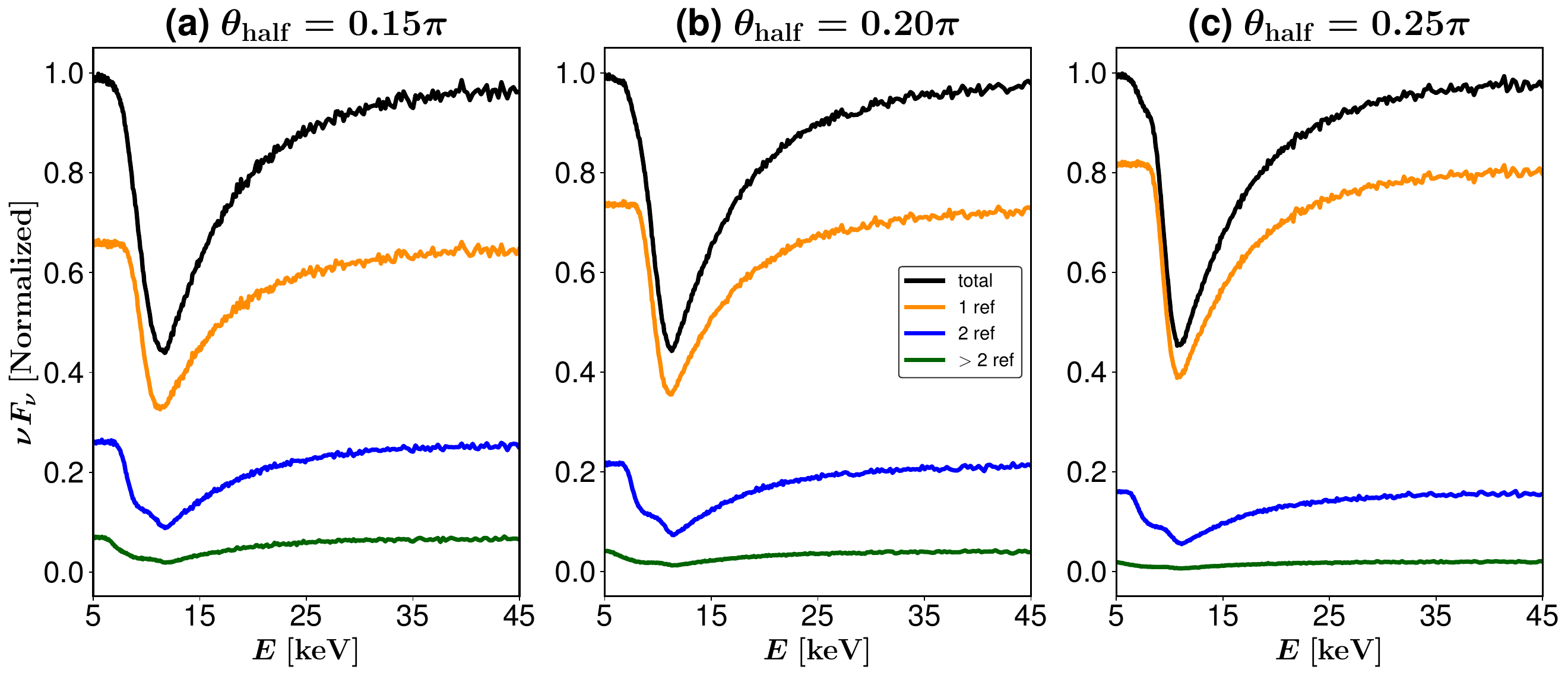}
    \caption{The dependence of the absorption edge profile on the half-open angle $theta_{\rm half}$. The color scheme and line styles are the same as in Fig.~\ref{cont_vinf}. Panel (b) is the result of the default set of model parameters, while panel (a) and (c) have different $\theta_{\rm half}$. As the funnel becomes narrower, the absorption edge becomes less skewed.}
    \label{cont_open}
\end{figure*}

\noindent We also investigate the Fe K absorption edge. The zoomed-in spectra of the absorption edge (after taking out the Fe line photons) are plotted in Fig.~\ref{cont_vinf}-\ref{cont_open}, for which the same parameter sets as in Fig.~\ref{vinf}- \ref{open} are used.  These spectra are also decomposed into different components corresponding to the order of reflection.

One can see from Fig.~\ref{cont_vinf} that the speed of the wind can impact the shape of the absorption edge. In the rest frame of the gas, the absorption edge has a skewed shape with a sharp drop at 8.8\,keV and a slow rise above this energy. However, due to the strong Doppler effect in the fast wind, the trough is blueshifted, and the dip starts at lower energy, which together makes the absorption feature shallower and more symmetric in shape. A similar effect occurs if the wind has a faster acceleration as shown in Fig.~\ref{cont_Racc}, although the impact of $R_{\rm acc}$ in the absorption profile is mild.

Moreover, it can be noticed that including multiple reflections further manifests the blueshift of the trough and makes the absorption edge start at an even lower energy. If $H_{\rm LP}$ decreases or $\theta_{\rm half}$ decreases, it will be less likely for photons to escape after the first reflection, and the beaming effect of the photons from the first reflection will be reduced, both of which promote more multiple reflections to happen in the lower funnel region. Therefore, as shown in Fig.~\ref{cont_hlp} and Fig.~\ref{cont_open}, a lower $H_{\rm LP}$ or a smaller $\theta_{\rm half}$ both produce less skewed absorption edge profiles.

In summary, the observed absorption edge profile can become more symmetric when the wind moves faster, or the corona has a lower height, or the super-Eddington funnel is narrower. The Fe line and the absorption edge profile can therefore be combined together to more effectively constrain the wind, funnel, and corona properties in super-Eddington accretion flows.


\section{Consideration of different ionization states of Fe}\label{FeIonization}

\begin{figure}[!htbp]
    \centering
    \includegraphics[width=0.5\linewidth]{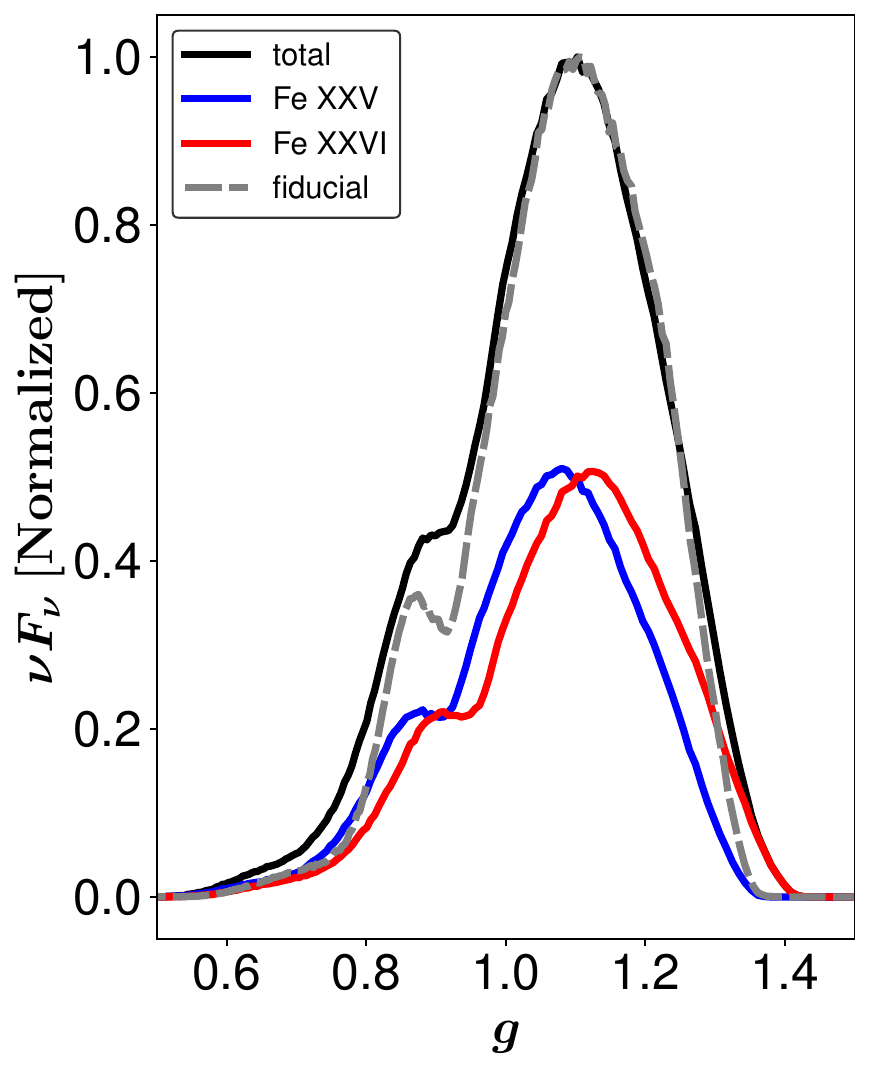}
    \caption{Line profiles at different ionization states. The x-axis is defined as $g = E_{\rm obs}/(6.7\ {\rm keV})$. The y-axis is the normalized flux. The grey dashed line is the total line flux as in Fig.~\ref{FeLine}~(b) with 100\% contribution from the Fe K$\alpha$ line at 6.7 keV. The black black line is the total line flux, which comes from a combination of 50\% Fe K$\alpha$ line at 6.7 keV (blue solid line) and 50\% Fe K$\alpha$ line at 6.97 keV (red solid line) in terms of photon number.}
    \label{H_He}
\end{figure}

\noindent In this study we have focused on the He-like Fe K$\alpha$ line with an intrinsic energy of 6.7 keV. In practice Fe can have different ionization states and complex ionization structures. For example, H-like Fe line at 6.97 keV is also likely produced when the ionization state is high such as in super-Eddington accretion flows. We therefore consider the extreme case that Fe K$\alpha$ lines of 6.7 keV and 6.97 keV are both produced with equal photon number contributions. We show in Fig.~\ref{H_He} the total line profile for this case with  Compton scattering included. One can see that the two Fe K$\alpha$ line structures can merge together due to their small energy difference. The double-peak feature is still produced, since the energy difference between the two Fe lines is not as large as the gap between the two peaks of each line. However, the two peaks of the merged line profiles are not as distinct as those for each individual line.

\restartappendixnumbering

\end{CJK*}
\end{document}